\documentclass[useAMS,usenatbib]{mn2e}
\usepackage{graphicx}
\usepackage{natbib}
\usepackage{color}

\newcommand{\Msun}{M_{\sun}}
\newcommand{\hi}{H\,{\sc i}}
\newcommand{\kms}{km~s$^{-1}$}
\newcommand{\rone}{R\hbox{$^{\prime}_{1}$}}  %perpendicular to bar
\newcommand{\amax}{\ensuremath{a_{\epsilon}}}

\newcommand{\afour}{\ensuremath{A_{4}}}
\newcommand{\bfour}{\ensuremath{B_{4}}}
\newcommand{\lbar}{\ensuremath{L_{\mathrm{bar}}}}
\newcommand{\dpa}{\ensuremath{\Delta \mathrm{PA}}}

\newcommand{\rbox}{\ensuremath{R_{\rm box}}}

\title[B/P Bulges in Moderately Inclined Galaxies]{Peanuts at an Angle: Detecting and
Measuring the Three-Dimensional Structure of Bars in Moderately Inclined Galaxies}

\author[P. Erwin \& V.P. Debattista]{Peter Erwin$^{1,2}$\thanks{E-mail: 
erwin@mpe.mpg.de} and Victor P. Debattista$^{3}$ \\
$^{1}$Max-Planck-Insitut f\"{u}r extraterrestrische Physik, Giessenbachstrasse, 85748 Garching, Germany \\
$^{2}$Universit\"{a}ts-Sternwarte M\"{u}nchen, Scheinerstrasse 1, D-81679 M\"{u}nchen, Germany \\
$^{3}$Jeremiah Horrocks Institute, University of Central Lancashire, Preston PR1 2HE, UK}

\begin{document}

\maketitle

\label{firstpage}

% The Abstract:
\begin{abstract}
We show that direct detection and measurement of the vertically thickened parts of bars
(so-called ``boxy'' or ``peanut-shaped'' bulges) is possible not only for
edge-on galaxies but also for galaxies with moderate inclinations ($i <
70\degr$), and that examples are relatively common in the nearby Universe.  

Analysis of a sample of 78 nearby, moderately inclined ($i \la 65\degr$)
early-type (S0--Sb) barred galaxies shows that the isophotal signature of the
box/peanut can usually be detected for inclinations as low as $i \sim 40\degr$
-- and in exceptional cases down to $i \sim 30\degr$.  In agreement with the
predictions from $N$-body simulations, the signature is most easily detectable
when the bar's position angle is within $\sim 50\degr$ of the galaxy major axis;
in particular, galaxies where the bar lies very close to the minor axis do not
show the signature clearly or at all. For galaxies with $i = 40$--65\degr{} and
relative angles $< 45\degr$, we find evidence for the signature $\approx 2/3$ of
the time; the true frequency of box/peanut structures in bars may be higher.

Comparison with $N$-body models also allows us to link observed
photometric morphology with 3D physical structures, and thus estimate
the relative sizes of box/peanut structures and bars. For our local
sample, we find that box/peanut structures range in radial size
(measured along the bar major axis) from 0.4--3.8 kpc (mean $= 1.5 \pm
0.9$ kpc) and span 0.26--0.58 of the bar length (mean of $0.38 \pm
0.08$). This is a clear observational confirmation that when bars
thicken, it is not the entire bar which does so, but only the inner part.

This technique can also be used to identify galaxies with bars which have
\textit{not} vertically thickened. We suggest that NGC~3049 and IC~676 may be
particularly good examples, and that the fraction of S0--Sb bars which
\textit{lack} box/peanut structures is at least $\sim 13$\%.

\end{abstract}

% Keywords:

\section{Introduction}

For a long time, the vertically thickened inner regions of disc galaxies have
been referred to as ``bulges'', for straightforward descriptive reasons. For
almost as long, these have been understood to be spheroidal, kinematically hot
structures, akin to elliptical galaxies. However, peculiar exceptions have also
been known for some time -- in particular, cases where bulges seen in edge-on
galaxies have a distinctly ``boxy'' or even ``peanut-shaped'' morphology. A
series of imaging studies
\citep{jarvis86,de-souza87,shaw87,dettmar90,lutticke00a} gradually demonstrated
that such structures are actually quite common; L\"utticke et al.\ found that
$\sim 45$\% of edge-on bulges in S0--Sd galaxies are boxy or peanut-shaped. Even
the Galaxy's own bulge has turned out to be boxy \citep[e.g.,][]{kent91,dwek95}.
The peculiarity  is not just morphological: several early stellar-kinematic
studies noted that strongly boxy or peanut-shaped bulges exhibited
\textit{cylindrical} stellar rotation
\citep[e.g.,][]{bertola77,kormendy-illingworth82}, something not at all
characteristic of elliptical galaxies.

Although several models have been proposed for boxy or peanut-shaped bulges,
such as their being the results of minor mergers \citep[e.g.,][]{binney85}, the
most successful explanation has come from investigations of bar formation and
evolution. A pioneering 3D $N$-body study by \citet{combes81} noted that the
bars which formed in their simulation showed ``a peanut-shape morphology'' when
the model was viewed edge-on with the bar perpendicular to the line of sight, 
an appearance similar to classic peanut-shaped bulges in systems such as
NGC~128. In the early 1990s, simulations of galaxy discs clearly showed that a
vertically unstable ``buckling'' phase often followed the formation of a bar
\citep[e.g.,][]{combes90,raha91}; the morphology and cylindrical kinematics of
the resulting structure matched observations of boxy and peanut-shaped bulges
(see \citealt{athanassoula05} and \citealt{debattista06} for reviews). This
rapid, asymmetric buckling phase is usually assumed to be driven by a global
bending instability \citep[e.g.,][]{merritt94}. However, alternate formation
mechanisms which involve the resonant heating or trapping of stellar orbits have
been suggested \citep{combes90,quillen02,debattista06}

Other theoretical studies have investigated the underlying orbital structure
which may support this morphology
\citep[e.g.,][]{pfenniger85,pfenniger91,patsis02b,martinez-valpuesta06}, explored conditions under
which it may be promoted or suppressed
\citep[e.g.,][]{berentzen98,athanassoula02,athanassoula05,debattista06,wozniak09}, 
and even suggested that multiple phases of buckling and vertical growth can
take place \citep{athanassoula05b,martinez-valpuesta06}.

Evidence confirming the association of bars with boxy/peanut-shaped
(B/P) bulges in real galaxies has come primarily from spectroscopy of
edge-on galaxies. The major-axis kinematics of ionized gas
\citep{kuijken95,merrifield99,bureau99a,veilleux99} and stars
\citep{chung04} in edge-on galaxies with boxy or peanut-shaped bulges
displays the characteristic imprint of bars, as predicted by orbital
analyses and simulations, both pure $N$-body
\citep{athanassoula99,bureau05} and hydrodynamical
\citep[e.g.,][]{athanassoula99}.  (Note, however, that the appearance of
this feature in gas kinematics requires that the so-called $x_{2}$ orbit
family be present, which requires that the bar have an inner Lindblad
resonance, something not all bars necessarily have.) In addition,
near-IR imaging of edge-on systems indicates that B/P bulges are
accompanied by larger-scale extensions in the disc of the galaxy,
suggestive of the vertically thin outer zones of bars
\citep{lutticke00b,bureau06}. The frequency of boxy and peanut-shaped
bulges is consistent with most barred galaxies having vertically
thickened inner regions \citep{lutticke00a}.

Finding B/P bulges is relatively easy in edge-on galaxies -- provided the
features are strong and not overwhelmed by a large classical bulge, \textit{and}
that the bar is favorably aligned: i.e., close to perpendicular to the line of sight.
(As the bar orientation shifts closer to end-on, the projection of the B/P bulge
becomes rounder and thus harder to distinguish from a classical bulge.) However,
measuring the characteristics of the rest of the bar -- its length, orientation,
strength, shape, etc.\  -- is much more difficult, both due to dust extinction
and to the superposition of stellar light from various regions of the disc along
the line of sight.  This same difficulty in identifying and measuring the
``flat'', planar parts of bars also makes it difficult to  find examples of
galaxies with bars which have \textit{not} buckled. It would clearly be useful
if there were a way to identify the B/P structure in face-on bars, or even in
bars of moderately inclined galaxies, where the in-plane structure of the bar
and disc is still discernable.

One promising approach is the direct detection of stellar-kinematical features
associated with B/P bulges in less inclined galaxies, as proposed by
\citet{debattista05}. \citet{mendez-abreu08} demonstrated that this is possible
by detecting the kinematic signature of a B/P structure in the low-inclination
($i = 26\degr$) barred galaxy NGC~98.  However, this method is most useful when
the galaxy has a very low inclination ($i < 30\degr$), and it requires high-S/N
spectroscopy and expensive allocations of telescope time (e.g., $\sim 3$h on
8--10m-class telescopes).

The standard approach for identifying B/P structures from imagery has been to
look at very highly inclined or edge-on galaxies; minimum inclinations of $\sim
75\degr$ or 80\degr{} have been suggested \citep[e.g.,][]{jarvis86,shaw90}.
There \textit{have} been isolated reports of B/P structures in images of
galaxies which are highly inclined but not actually edge-on (i.e., inclinations
$\sim 70$--85\degr). \citet{buta90} noted the peculiar ``inner hexagonal zone''
of NGC~7020 ($i = 69\degr$), even going so far as to suggest a possible
connection with box/peanut-shaped structures from the simulations of
\citet{combes81}. A few years later, \citet{bettoni94} pointed out the case of
NGC~4442 ($i = 72\degr$), which they explicitly identified as hosting a
thickened bar with projected isophotes similar to those of the B/P structure in
the $N$-body simulations of \citet{combes90}; they also found evidence for
cylindrical rotation in the stellar kinematics, similar to that seen in the
simulations. Likewise, \citet{quillen97} identified the bar of NGC~7582 ($i \sim
70\degr$) as hosting a peanut-shaped bulge.

More recently, \citet{athanassoula06} used a relatively deep 2MASS image
of M31 ($i = 77\degr$) to show that it, too, has a boxy bulge embedded
within a longer bar \citep[see also][]{beaton07}.  By comparing 
isophotes and surface-brightness profiles from cuts parallel to
the major axis of the galaxy with isodensity contours and
parallel cuts from a selection of $N$-body simulations, they
demonstrated that the morphology of M31 immediately outside its
classical bulge was consistent with that of a bar having both a B/P
structure and an outer, flatter region seen at high inclination and a
slight offset with respect to the galaxy's major axis. (They also
found similarities between M31's  gas kinematics and predictions from
gas flow in barred-galaxy simulations.)

In this paper, we demonstrate that there is a consistent set of isophotal
features which makes identification of B/P bulges in images of
\textit{moderately} inclined ($i < 70\degr$) galaxies quite possible, and that
numerous examples of galaxies with these features exist. We find that B/P
structures can be identified in images even when the inclination is as low as $i \sim
30\degr$.

Throughout this paper, we assume a Hubble constant of $H_{0} = 72$ km s$^{-1}$ Mpc$^{-1}$.

\subsection{A Note on Terminology}

In the course of this study, we use the words ``box'' and ``boxy'' rather often,
in reference to several related  phenomena. For clarity, we will use the terms
\textbf{boxy/peanut-shaped bulge} and \textbf{B/P structure} to refer to a
specific 3D stellar structure: the vertically thickened inner part of a bar, as
discussed above. We will also use the terms \textbf{boxy bar} and
\textbf{box+spurs}: these refer to a \textit{2D} morphological feature seen in
the isophotes of moderately inclined galaxies. Much of this paper is devoted to
demonstrating that the existence of the former (3D) structure  explains the
presence of the latter (2D) phenomenon in real galaxies.

% Figure 1
\begin{figure}
\begin{center}
\includegraphics[scale=0.62]{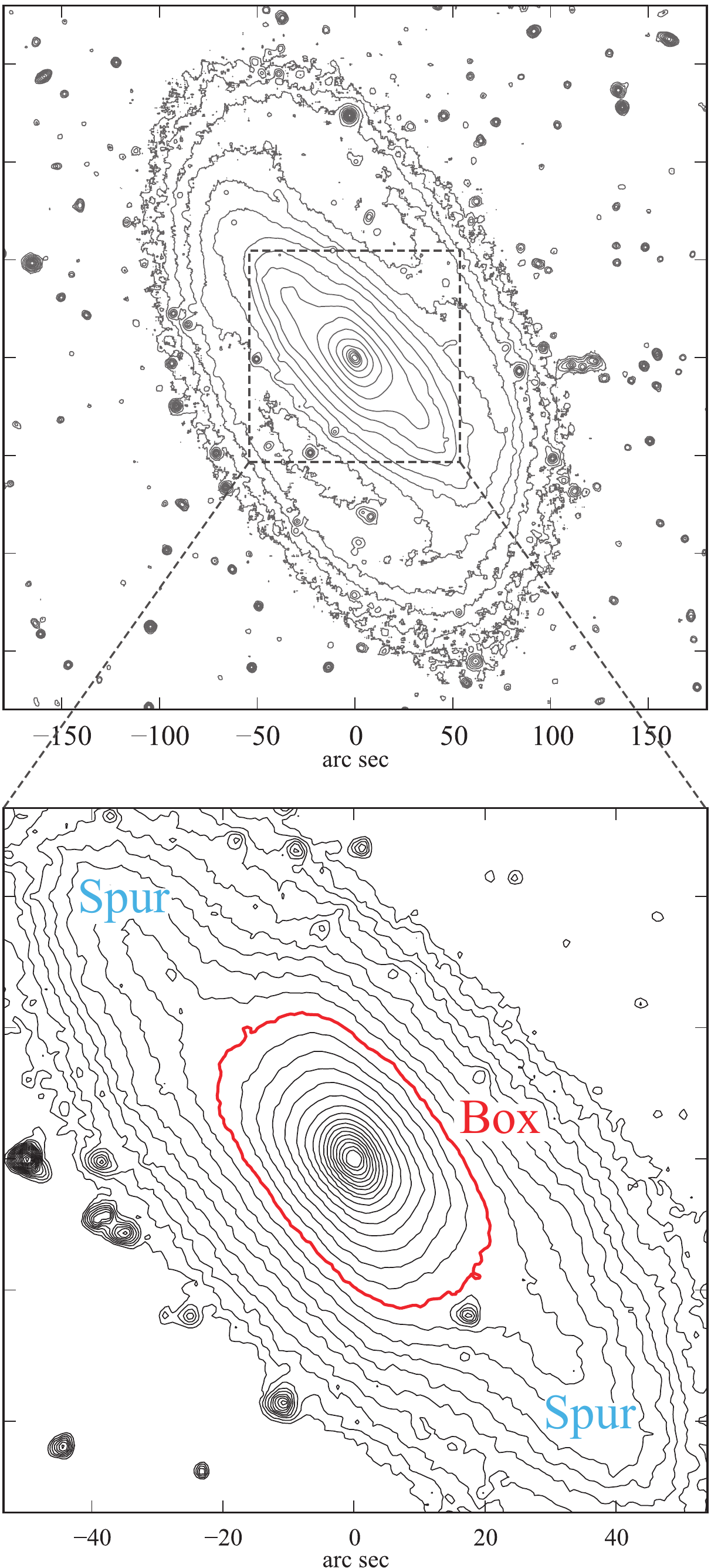}
\end{center}

\caption{Logarithmically scaled isophotes of the barred Sa galaxy NGC~5377,
showing the ``boxy-bar''/``box+spurs'' morphology.  \textbf{Top panel:} $R$-band
isophotes from \citet{erwin03}; \textbf{bottom panel:} close-up of the bar,
showing isophotes from an archival \textit{Spitzer} IRAC1 (3.6$\mu$m) image. The
broad, nearly rectangular region in the inner part of the bar is the ``box'',
labelled in red; the narrow ``spurs'' projecting outside make up the outer part
of the bar and are labelled in blue. In both panels N is up and E is to the
left. \label{fig:box+spurs}}

\end{figure}

\section{The Visual Phenomenon: Examples of Boxy Bars from Various Sources}\label{sec:boxy-bars}

We begin by discussing a peculiar morphology seen in some moderately inclined
galaxies. Figure~\ref{fig:box+spurs} shows a characteristic example: the SBa
galaxy NGC~5377, which has a strong bar with a radial size of $\sim 65\arcsec$
and a position angle on the sky of 45\degr. The inner part of the bar ($r <
30\arcsec$) has isophotes which are rather broad and distinctly ``boxy'' in
shape -- in this particular case, almost rectangular. At larger radii, the
isophotes of the bar appear as narrower projections outside the boxy zone; we
term these projections ``\textbf{spurs}''. As we will discuss below, these
narrower projections are almost always slightly offset from the major axis of
the inner isophotes. We call this composite phenomenon the
``\textbf{box+spurs}'' or ``\textbf{boxy-bar}'' morphology; three more examples
can be seen in Figure~\ref{fig:offset-spurs}. \citet{laurikainen11} recently
noted the presence of bars with ``boxy/peanut/x-shaped structures (B$_{x}$)'' in
a handful of moderately inclined galaxies, including NGC~5377; this is
undoubtedly the same thing.

A list of nearby galaxies showing this morphology is given in
Table~\ref{tab:galaxies}. This list -- which is not meant to be comprehensive or
complete -- is based on inspection of a variety of data sources, including the
Sloan Digital Sky Survey \citep[Data Release 7;][]{york00,sdss-dr7}, as well as
images available via NED, including those from \citet{mh01}, \citet{eskridge02},
and \citet{knapen03}. We also identified some candidates by examining published
isophotal maps, including those of \citet{jung97}, \citet{peletier99}, and
\citet{rest01}. Three galaxies not already in our list were added from the set
of non-edge-on galaxies with ``B$_{x}$'' classifications in
\citet{laurikainen11}.\footnote{We exclude NGC~2549, NGC~4220, NGC~5353, and
NGC~7332, since they are edge-on, or nearly so.}

\begin{table*}
\begin{minipage}{126mm}
    \caption{Galaxies with Boxy-Bar Signatures}
    \label{tab:galaxies}
    \begin{tabular}{lllllrlll}
\hline
Name     & Type     & Distance & Source &  $M_{B}$  & $i$  & Offset Spurs & Lead/Trail & Source \\
(1)      & (2)      & (3)      & (4)    &  (5)      & (6)  & (7)        & (8)        & (9)    \\
\hline
 M31     &                SA(s)b              &  0.79 & 1 & $-21.20$ &   77 &  Yes & trail  & 1\\ % rotation = CCW
NGC 1023 &    SB(rs)$0^{-}$                   &  11.1 & 3 & $-20.94$ &   69 &  Yes & --- & 2\\ 
NGC 1079 &  (R$_{1}$R$^{\prime}_{2})$SAB(r$^{\prime}$l)a   &  17.2 & 4 & $-19.11$ &   53 &  Yes & trail & 2\\ % rot = CCW
NGC 1350 &        \rone{}SB(r)ab              &  16.7 & 2 & $-20.36$ &   57 &  Yes & trail & 2\\    % rot = CW
NGC 1375 &        SAB$0^{0}$                  &  31.5 & 3 & $-19.39$ &   71 &  Yes & ---  & 2\\ 
\textbf{NGC 1415} &     (R)SAB0/a(s)          &  19.2 & 4 & $-19.13$ &   65 &  Yes & lead & 3\\ 
NGC 1784 &  SB(r)c                            &  30.5 & 2 & $-21.15$ &   52 &  Yes & lead & 2\\    % rot = CW
\textbf{NGC 1808} &        (R)SAB(s)a         &  12.3 & 2 & $-20.17$ &   50 &  Yes & lead & 2\\  % Yes
\textbf{NGC 2442} &        SAB(s)bc pec       &  16.2 & 4 & $-20.78$ &   62 &  Yes & lead & 2\\ 
\textbf{NGC 3185} &                (R)SB(r)a  &  17.5 & 4 & $-18.61$ &   49 &  Yes & lead  & 2\\  % Yes
NGC 3595 &             E?                     &  33.6 & 4 & $-19.89$ &   64 &  Yes & --- & 2\\ 
\textbf{NGC 3627} &               SAB(s)b     &  10.1 & 5 & $-20.92$ &   65 &  Yes & trail & 2\\ 
NGC 3673 &           SB(rs)b                  &  17.4 & 2 & $-19.24$ &   42 &  Yes & lead & 2\\ 
NGC 3885 &           SA(s)0/a                 &  23.4 & 4 & $-19.57$ &   67 &  Yes & lead & 2\\    % rot = CW [prob.]
NGC 3992 &           SB(rs)bc                 &  22.9 & 2 & $-22.38$ &   56 &  Yes & trail & 2\\    % rot = CCW
\textbf{NGC 4123} &           SB(r)c          &  14.9 & 2 & $-19.26$ &   45 &  Yes & lead & 2\\ 
\textbf{NGC 4192} &          SAB(s)ab         &  13.6 & 2 & $-20.74$ &   79 &  Yes & trail & 2\\ 
\textbf{NGC 4293} &       (R)SB(s)0/a         &  16.5 & 6 & $-20.35$ &   63 &  Yes & trail & 4\\   % rot = CW [prob.]
NGC 4429 &           SA(r)$0^{+}$             &  16.5 & 6 & $-20.20$ &   62 &  Yes & --- & 2,4\\ 
NGC 4442 &             SB(s)$0^{0}$           &  15.3 & 7 & $-19.64$ &   72 &  yes? & --- & 5\\ 
NGC 4462 &           SB(r)ab                  &  23.9 & 4 & $-20.07$ &   71 &  Yes & trail? & 2\\ 
NGC 4535 &           SAB(s)c                  &  15.8 & 5 & $-20.62$ &   44 &  Yes & lead & 2\\  % rot = CCW
\textbf{NGC 4725} &             SAB(r)ab pec  &  12.4 & 5 & $-20.69$ &   42 &  Yes & trail & 2\\  % Yes
\textbf{NGC 5377} &                (R)SB(s)a  &  27.1 & 4 & $-20.29$ &   59 &  Yes & trail & 2\\  % Yes
NGC 5448 &                (R)SAB(r)a          &  31.5 & 4 & $-20.75$ &   66 &  Yes & lead & 4\\   % rotation = CW
\textbf{NGC 5641} &  (R$^{\prime}$)SAB(r)ab   &  70.2 & 8 & $-21.48$ &   58 &  Yes & trail & 2\\  % rotation = CW
NGC 6384 &               SAB(r)bc             &  25.9 & 2 & $-21.52$ &   47 &  Yes & lead  & 2\\
NGC 7020 &       (R)SA(r)0$^{+}$              &  40.5 & 4 & $-20.56$ &   69 &  Yes & ---  & 6\\ 
NGC 7582 &        \rone{}SB(s)ab              &  23.0 & 2 & $-20.94$ &   68 &  No? & ---  & 7\\     % rot = CW
\textbf{IC 5240}  &        SB(r)a             &  21.8 & 4 & $-19.23$ &   49 &  Yes & lead  & 8\\ 
ESO 443-39  &       S0?                       &  40.3 & 4 & $-19.55$ &   57 &  Yes & ---  & 2\\ 
UGC 3576 &      SB(s)b                        &  85.0 & 4 & $-20.87$ &   60 &  Yes & trail  & 2\\ 
UGC 11355 &         Sb                        &  63.1 & 4 & $-20.32$ &   58 &  Yes & ---  & 2\\ 
\hline
\end{tabular}

\medskip

A list of galaxies containing boxy-bar features. This list is not intended to
be complete or comprehensive; see Table~\ref{tab:wiyn} for examples in a
well-defined local sample. (1) Galaxy name; boldface type indicates a
particularly strong/emblematic example of the boxy-bar/box+spurs morphology. (2)
Hubble type from RC3. (3) Distance in Mpc. (4) Source of distance: 1 = mean of
distances in NED; 2 = Tully-Fisher distance from \citet{tully09}; 3 = SBF
distance from \citet{tonry01}, including metallicity correction from
\citet{mei05}; 4 = HyperLeda redshift (corrected for Virgo-centric infall); 5 =
Cepheid distance from \citet{freedman01}; 6 = mean Virgo Cluster distance from
\citet{mei07}; 7 = SBF distance from \citet{blakeslee09}; 8 = Tully-Fisher
distance from \citet{willick97}. (5) Absolute $B$ magnitude, from HyperLeda
$B_{tc}$ and our adopted distance. (6) Inclination. (7) Indicates whether spurs
extending out of boxy zone are offset from major axis of boxy interior. (8)
Indicates whether offset spurs, if present, lead or trail (assuming main spiral
pattern, if it exists, is trailing). (9) Source of identification: 1 =
\citet{athanassoula06}; 2 = this paper; 3 = \citet{garcia-barreto00}; 4 =
\citet{laurikainen11}; 5 = \citet{bettoni94}; 6 = \citet{buta90}; 7 =
\citet{quillen97}; 8 = R. Buta, private comm.
   
\end{minipage}
\end{table*}

\subsection{Spurs: Leading and Trailing}\label{sec:box+spurs}

\begin{figure*}
\begin{center}
\includegraphics[scale=0.75]{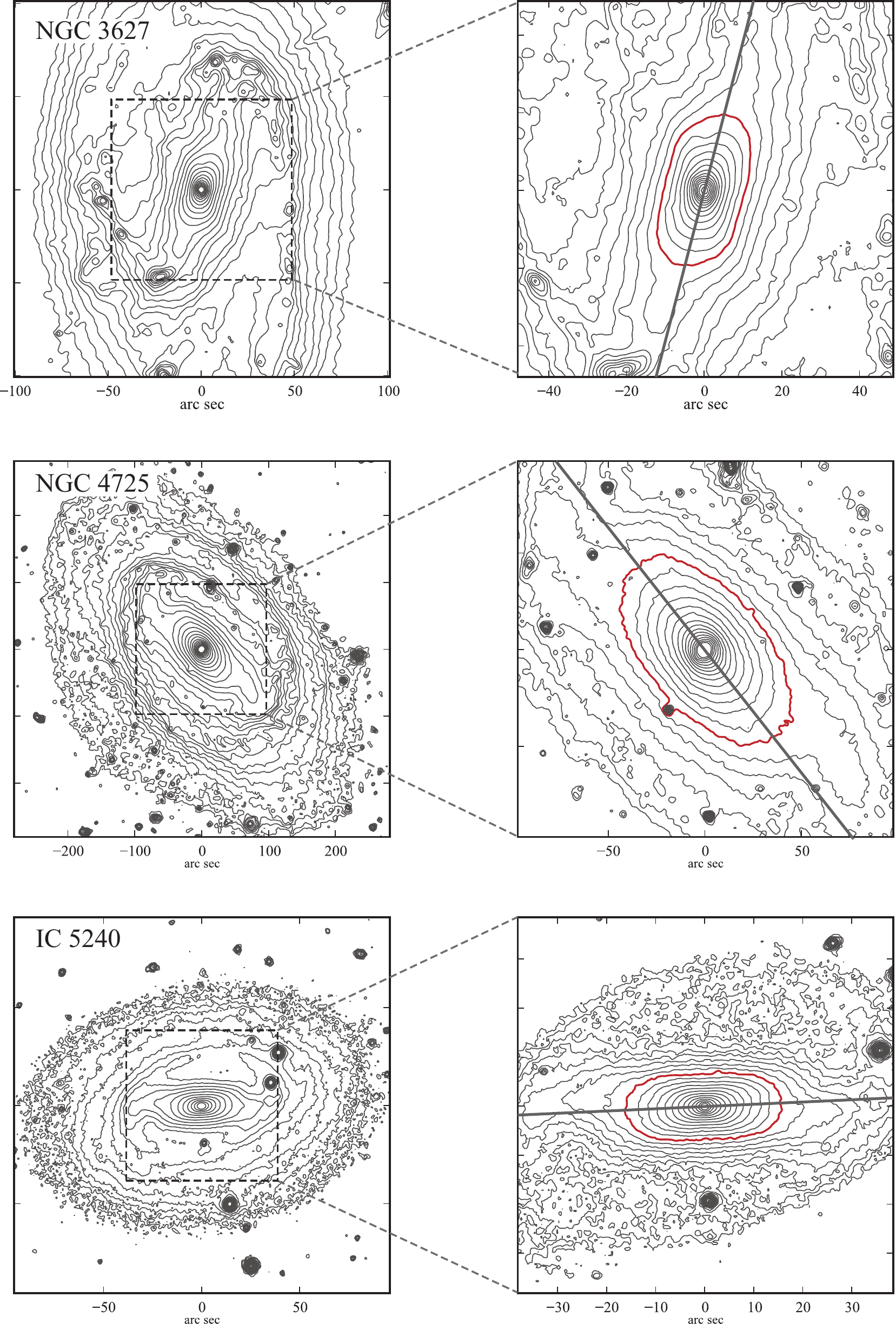}
\end{center}

\caption{Three more examples of box+spurs morphology in barred galaxies,
emphasizing the offset nature of the spurs: NGC~3627 \citep[top, \textit{Spitzer} IRAC1
image from the SINGS project;][]{kennicutt03}, NGC~4725 (same image source), and
IC~5240 (bottom; left-hand panel is $R$-band image from \citet{koopmann06},
right-hand panel is $K$-band image from \citet{mrk97}).  Dark grey lines in
right-hand panels indicate approximate position angle of the boxy regions
(outlined with thicker red contour lines); note that the spurs just outside this
region are displaced with respect to the lines. Comparison with spiral arms
indicates that the spurs are ``trailing-edge'' in the case of NGC~3627 and
NGC~4725, but ``leading-edge'' in IC~5240. N is up and E is to the left in all
panels.\label{fig:offset-spurs}}

\end{figure*}

One of the most striking aspects of the boxy-bar morphology is the existence of
the narrow spurs extending beyond the broader boxy region. These spurs are
usually \textit{offset} with respect to the major axis of the interior
isophotes. This can be seen in Figures~\ref{fig:box+spurs} and
\ref{fig:offset-spurs}. The diagonal grey lines in Figure~\ref{fig:offset-spurs}
indicate the major axis of the boxy regions, where the isophotes have the
general shape of rounded rectangles; the narrow spurs extending outside the boxy
region are symmetrically offset from this axis.

Could the offset spurs be just an illusion produced by dust?  We know that bars
often have strong dust lanes running along the leading edges of the bar
\citep[e.g.,][]{athanassoula92b}, so in principle offset spurs could be the
result of extinction along the bar leading edges.  In that case, however, we
would expect to see only \textit{trailing}-edge spurs, whereas in reality we
see both. In fact, we see them with approximately equal frequency:
Table~\ref{tab:galaxies} has 12 examples of trailing-edge spurs and 13 leading-edge
examples.  In addition, we see offset spurs in near-IR imaging (e.g.,
Figures~\ref{fig:box+spurs} and \ref{fig:offset-spurs}), where dust extinction
is weaker or absent, and in S0 galaxies with no detectable gas or dust (e.g.,
NGC~1023, NGC~3595, NGC~4442, NGC~4429, and ESO~443-39).

% Figure 2
% wide figure -- use figure*

Finally, as we will show below (Section~\ref{sec:n-body-box-spurs}), the
$N$-body models (which are by nature dust-free) have the appearance of spurs as
well, and they make a specific prediction about when the spurs will be offset,
and in which direction -- a prediction which the observed galaxies match quite well.

\section{$N$-body Models}

We compare observations with a number of collisionless $N$-body simulations.  We
rely mainly on three such simulations.  The first, run A, has not been previously
published.  It is a disc galaxy evolving in the prolate halo B described in
\citet{debattista08}. Briefly, this halo was produced by the merger of two
spherical haloes starting at rest, 800 kpc apart. The disc grown in this model
was exponential with a scale-length of 6 kpc, a Gaussian scale-height $z_{\rm d}
= 300$ pc, a mass of $7 \times 10^{10} \, \Msun$ and Toomre Q $= 1.5$.  The
initial disc is oriented with its angular momentum perpendicular to the short
axis of the (initial) halo, where it remains throughout the simulation.  The
simulation was evolved with {\sc pkdgrav}, as described in \citet{debattista08}.

The other two models, runs B and C, have already been published in
\citet{sellwood09}.  These are simulations which differ only in the seed of the
random number generator, which was used to set up the initial conditions.
Although the two simulations represent instances of the same system, they evolve
very differently as a result the physically very stochastic nature of disc
galaxies.  The bar amplitude evolution of these two models can be seen in
Figure~5 of \nocite{sellwood09}Sellwood \& Debattista: run C is largely growing
in strength throughout the simulation and ends in the cluster of lines at the
highest amplitude, while run B is strongly weakened by buckling before it starts
growing in strength again; this is the simulation with bar amplitude
intermediate between the strongest and weakest cases. Further details of these
simulations, including a discussion of the importance of stochastic effects, can
be found in \nocite{sellwood09}Sellwood \& Debattista.  For run B, we use
snapshots at $t = 200$ (in simulation units), which is shortly after the bar
forms but before it buckles, and at $t = 1000$ (at the end of the simulation,
after buckling and a period of bar growth).  For run C we use outputs at $t =
200$ (also before buckling) and at $t = 600$ (after the bar has recovered from
buckling, with a bar amplitude $A_{2} \sim 16$\% larger than for B at $t=1000$).

\subsection{The 3D Origins of Boxes and Spurs}\label{sec:n-body-box-spurs}

So what is the origin of the box+spurs morphology? Put simply, it is the result
of viewing a bar which has a vertically thick inner region (the B/P structure)
and a vertically thin outer region. At moderate to high inclinations, the
projection of the B/P structure forms the box, and the vertically thin outer bar
forms the spurs. This insight is a generalization of previous
work by \citet{bettoni94} and \citet{athanassoula06}, who compared projections
of highly inclined $N$-body models with single galaxies to come to similar 
conclusions.\footnote{In the case of M31, Athanassoula \& Beaton's ``elongations''
correspond to what we call spurs.}

\begin{figure*}
\begin{center}
\includegraphics[scale=0.7]{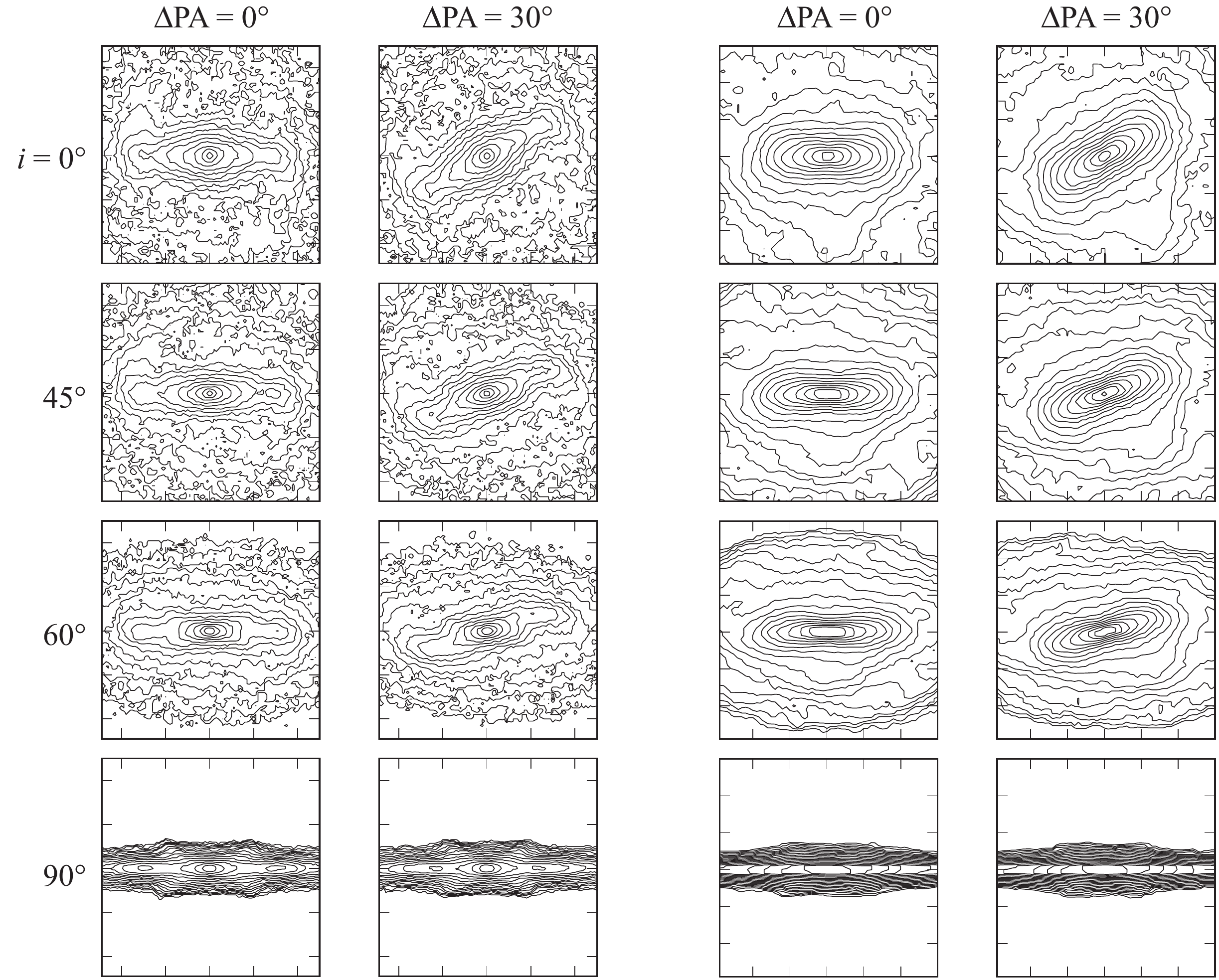}
\end{center}

\caption{Demonstration that B/P structures in $N$-body simulations produce the
boxy-bar (box+spurs) morphology. \textbf{Left panels:} log-scaled isodensity
contours of run A, viewed at different inclinations (face-on to
edge-on, top to bottom) and with different in-plane angles between bar and line
of nodes ($\dpa = {\rm PA}_{\rm bar} - {\rm PA}_{\rm disc}$, measured in the
galaxy plane); disc line of nodes is horizontal in all panels. When the simulation is
edge-on ($i = 90\degr$), the peanut-shaped bulge is visible; at lower
inclinations (60\degr{} and 45\degr), this projects to form the box of the
boxy-bar morphology, while the vertically thin outer part of the bar projects to
form the spurs. \textbf{Right panels:} same, but showing run B at
$t = 200$, where a B/P structure is \textit{not} present; consequently, no
box+spurs morphology is seen when the galaxy is moderately
inclined.\label{fig:n-body-basic}}

\end{figure*}

To demonstrate this, Figure~\ref{fig:n-body-basic} contrasts two different
$N$-body simulations of barred galaxies. On the left is run A, where
the bar has buckled and formed a distinct B/P structure, which can be seen in
the bottom left, edge-on panels. When seen at intermediate inclinations -- and
in particular, when seen with the bar at an intermediate angle \dpa{} with
respect to the line of nodes -- the box+spurs morphology emerges; this can be
seen most clearly in the $i = 60\degr$, $\dpa = 30\degr$ panel. In contrast, the
right-hand panels show a bar which has \textit{not} buckled. The views of this
simulation at moderate inclinations do not show a boxy-bar signature, even in
the most favorable $i = 60\degr$, $\dpa = 30\degr$ view.

% Figure 3

\begin{figure*}
\begin{center}
\includegraphics[scale=0.87]{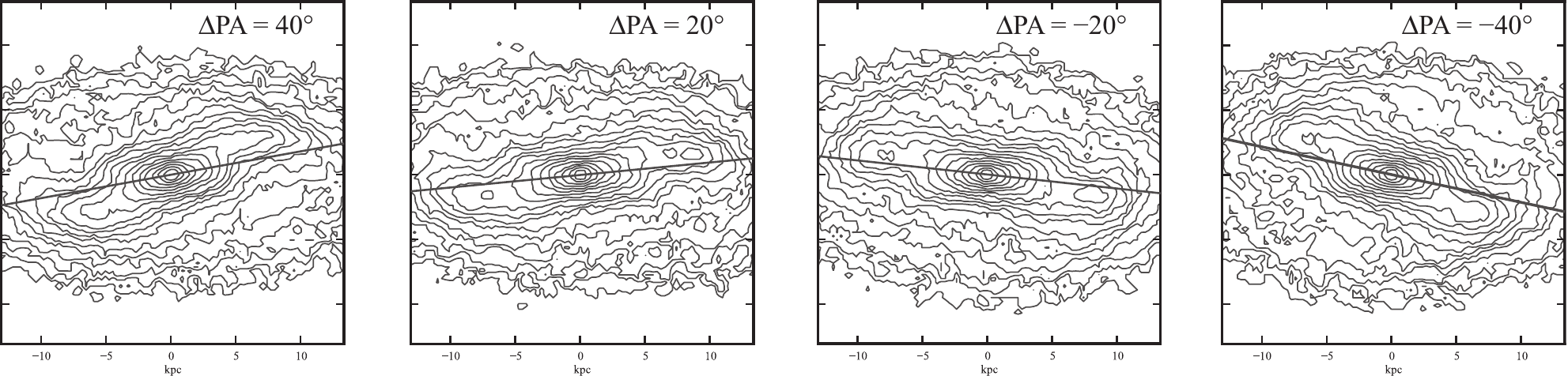}
\end{center}

\caption{$N$-body model A, viewed at $i = 60\degr$. Each panel shows a different
rotation of the bar relative to the line of nodes, which is horizontal in all
panels: from left to right, $\dpa = 40\degr$, 20\degr, $-20\degr$, and
$-40\degr$ (all measured in the galaxy plane). The dark grey diagonal lines
indicate the approximate position angle of the boxy regions; the spurs are
displaced with respect to these lines as in Figure~\ref{fig:offset-spurs}. The
sense of the displacement is always the same: rotated further away from the
line of nodes than the boxy region is.
\label{fig:n-body-offset-spurs}}

\end{figure*}

In Figure~\ref{fig:n-body-offset-spurs}, we can see that the \textit{offset}
nature of the spurs, pointed out in Section~\ref{sec:box+spurs} for real
galaxies, is in the following sense: the spurs are shifted \textit{away} from
the major axis, relative to the boxy inner zone. This is because the projection
of the B/P structure creates boxy-zone isophotes which are tilted closer to the
line of nodes than are the isophotes due to the projection of the outer, flat
part of the bar, which form the spurs. (Another way to view this is that when
inclination shrinks the observed angle between the outer, flat part of the bar
and the line of nodes, it shrinks the apparent angle between the inner part of
the bar and the line of nodes \textit{more}, making the boxy zone appear more
closely aligned with the line of nodes than the outer part of the bar.)

Is this consistent with what we see in real galaxies? Of the 35 galaxies where
we have been able to directly measure the position angle of the line of nodes
and the boxy region (see Table~\ref{tab:measurements} and the figures in the
Appendix), we find perfect agreement for all 24 galaxies where the position angles of
the line of nodes and the boxy region differ by $\ge 5\degr$. At smaller
relative position angles, we become vulnerable to errors in determining the line
of nodes, so that the sense of which direction the box is rotated relative to
the line of nodes becomes uncertain. None the less, for the eight galaxies in
which the relative angle between box and line of nodes is between 1\degr{} and
5\degr, six show the spurs offset in the correct direction. (Three more galaxies have
box and line of nodes position angles differing by $< 1 \degr$, making it effectively
impossible to determine the sense in which the box is rotated relative to the line
of nodes.)

% Figure 4

The correspondence between bar position angle and spur offset
shown by both simulated and real galaxies helps rule out other possible
explanations for the spurs. For example, in some galaxies the spurs
appear to blend smoothly into spiral arms which trail off the ends of
the bar -- e.g., NGC~3627 and NGC~4725 in Figure~\ref{fig:offset-spurs};
this might suggest that the spurs are somehow part of the spiral arms,
rather than the outer part of the bar. Inspection of
Figure~\ref{fig:n-body-offset-spurs} shows that projection effects
create the appearance of spiral twisting at the ends of the simulated
bars -- but the direction of the twist depends on the bar orientation,
so that the twisting is always \textit{towards} the line of nodes
(compare the far-left and far-right panels). Inspection of the galaxies
in Table~\ref{tab:galaxies}, along with galaxies from our local sample
with the boxy-bar morphology (Section~\ref{sec:local-sample}), reveals
nineteen galaxies where the spurs show signs of curvature; in sixteen of
these, the curvature is toward the line of nodes, which suggests that
this may indeed be an additional projection effect. Given that face-on
bars often (in both real galaxy and simulations) appear to have spiral
arms trailing off of the ends of bars, we should not be surprised to see
spurs blend into spiral arms at larger radii; this does not, however,
mean that the entirety of the spur \textit{is} a spiral
arm.\footnote{For NGC~3627 (Figure~\ref{fig:offset-spurs}), we
see curvature of the northern spur towards the major axis, which then gives way at
larger radii to a spiral arm twisting the \textit{opposte} direction,
which strongly argues that the spur is not an inward continuation of the
spiral arm.}

\begin{figure*}
\begin{center}
\includegraphics[scale=0.8]{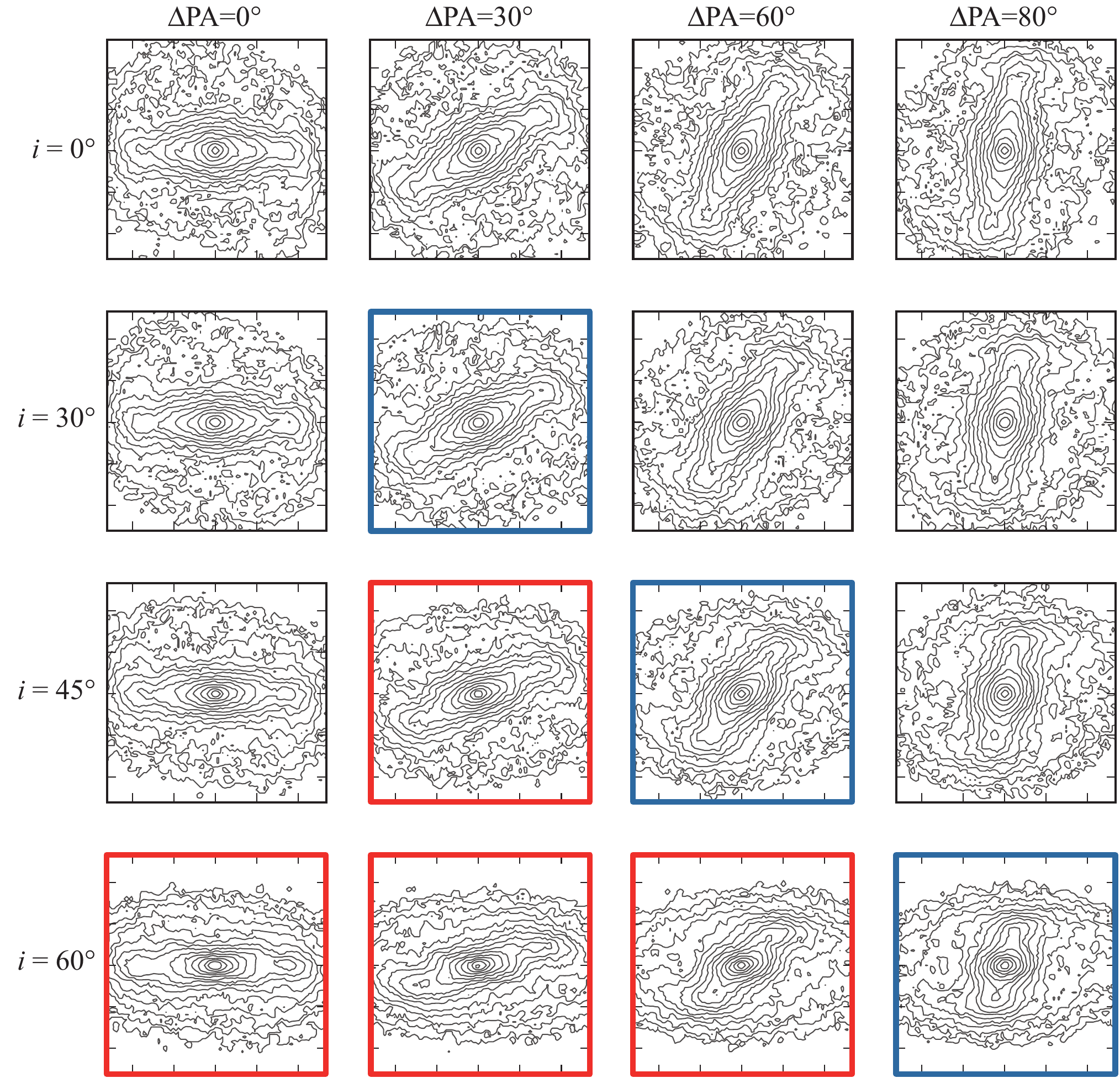}
\end{center}

\caption{Grid showing various projections of $N$-body model A. From left
to right, the projections show the bar rotated with respect to the line of nodes
(\dpa). The simulation is projected at increasing inclinations as one goes from
the top row (face-on) to the bottom ($i = 60\degr$); the disc line of nodes is
horizontal in all panels. Thick red lines outline panels where some version of
the box+spurs morphology is clearly visible; thick blue lines outline panels
where weaker versions of this morphology are (possibly)
visible.\label{fig:n-body-grid}}

\end{figure*}

Figure~\ref{fig:n-body-grid} shows a set of projections of one $N$-body
simulation, arranged by inclination and by the relative position angle of the
bar with respect to the line of nodes. We can clearly see that the box/peanut
structure becomes more visible as the inclination is increased, which is not
surprising (see also Figure~\ref{fig:n-body-basic}); what \textit{is} perhaps
surprising is that the signature of the projected box/peanut is visible when the
inclination is relatively low: it is clearly present for $i = 60\degr$, and also
present when $i = 45\degr$ and $\dpa$ is $\sim 30\degr$. In fact, we can see
weak traces of the signature in the $i = 30\degr$, $\dpa = 30\degr$ panel (i.e.,
the fact that the main axis of the outer part of the bar appears slightly offset
on opposite sides of the centre).

\begin{figure*}
\begin{center}
\includegraphics[scale=0.8]{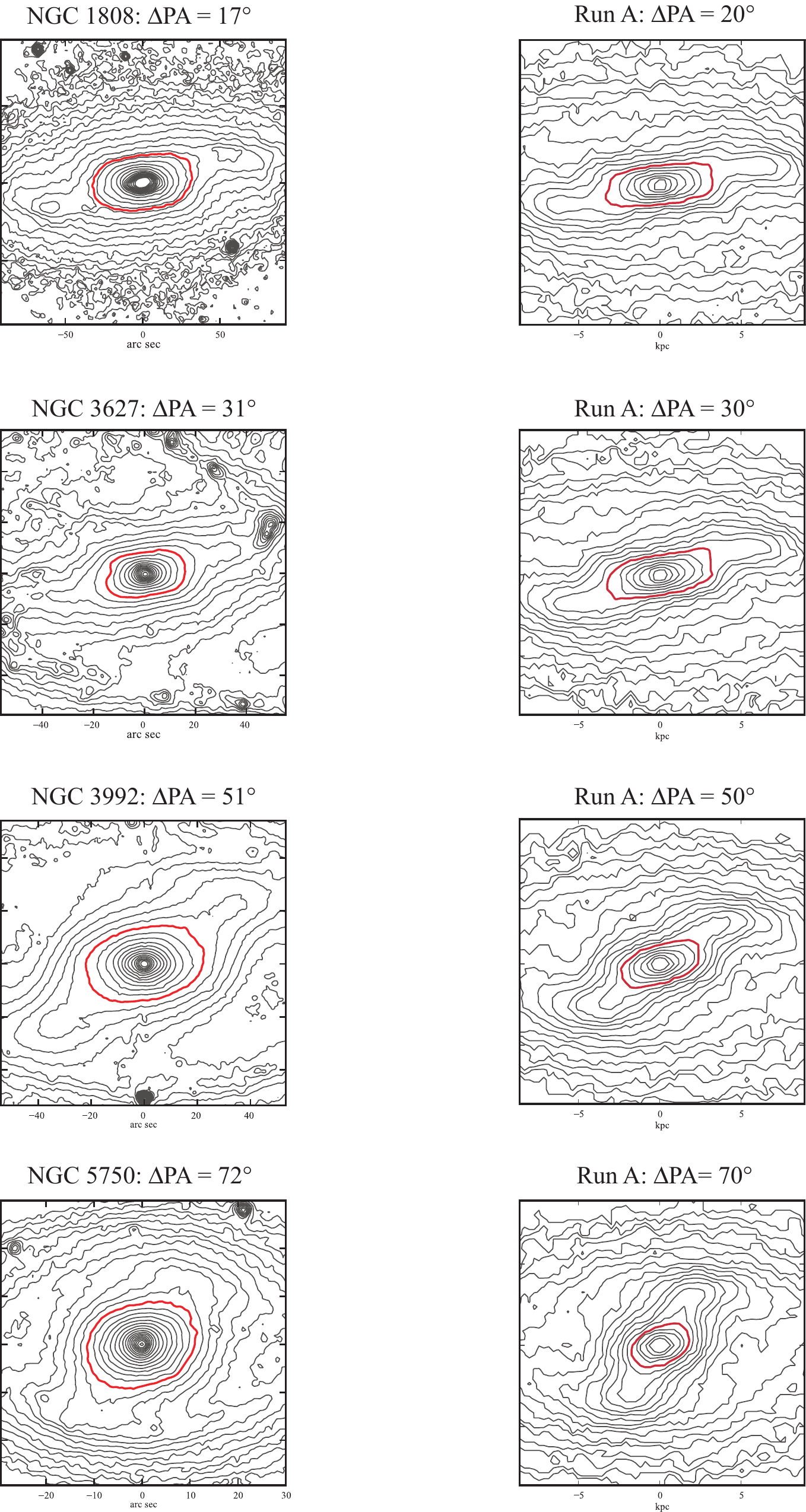}
\end{center}

\caption{Comparison of real galaxies and $N$-body models, and a demonstration of
how the offset spurs ``rotate'' from aligned with the major axis of the inner, boxy zone
(outlined in red) to projecting at an angle as the bar rotates further away from the
major axis. \textbf{Left panels}: Examples of real galaxies (all with
inclinations between 50\degr{} and 65\degr) where the (deprojected) angle \dpa{}
between the bar and the disc line of nodes increases from top to bottom. All
images have been rotated to make the disc major axis horizontal; the plots of
NGC~3627 and NGC~3992 has also been reflected about the vertical axis. (Images are
$H$-band for NGC 1808 and \textit{Spitzer} IRAC1 for the others.) \textbf{Right panels}:
Same, but now showing projections of $N$-body model A at $i
= 60\degr$.\label{fig:rotating-spurs}}

\end{figure*}

By looking at the full set of projections, we can also see that other, related
morphologies are indicators of projected box/peanut structures, even if they do
not match exactly the strong, paradigmatic form presented in
Section~\ref{sec:box+spurs}. For example, as $\dpa$ gets larger, we move from a
situation where the spurs appear to be parallel to the boxy zone to one
where the spurs appear to proceed from the \textit{corners} of the boxy zone at
some intermediate angle (e.g., the $i = 60\degr$ row of
Figure~\ref{fig:n-body-grid}, where this alternate morphology is clearly present
for $\dpa = 60\degr$.) This is shown more directly in
Figure~\ref{fig:rotating-spurs}, where we compare several real galaxies having
inclinations $\sim 50$--60\degr{} with projections of the same $N$-body
simulation.

Figure~\ref{fig:rotating-spurs} also illustrates how the basic features of the
box+spurs morphology in \textit{real} galaxies can be reproduced by $N$-body
models. Even though fine details may vary from galaxy to galaxy -- e.g., the
relative size of the boxy zone compared with the length of the spurs, the
apparent thickness of the spurs, etc. -- the same $N$-body model does an
impressive job of matching the basic isophote patterns in four different
galaxies.

\subsection{Matching Isophotal Features with 3D Stellar Structure in the Box/Peanut}\label{sec:3d}

It seems clear that we can identify the boxy zone in the box+spurs morphology with
the projection of the vertically thick B/P structure, and the spurs with the projection
of outer, vertically thin part of the bar. Can we quantify this? In particular: can we
devise a measurement of the boxy zone which corresponds to a measurement of the
3D B/P structure?

\begin{figure*}
\begin{center}
\includegraphics[scale=0.8]{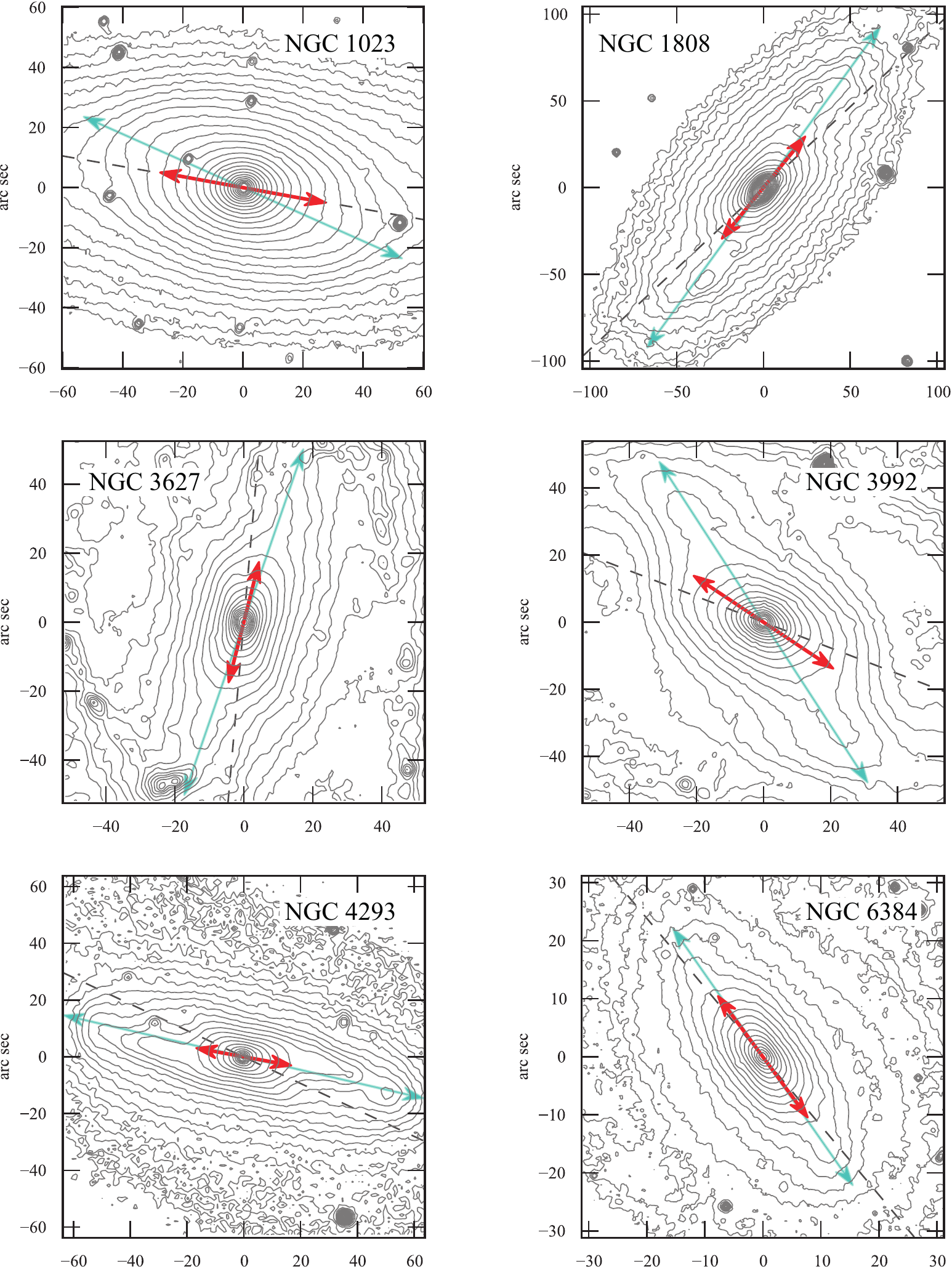}
\end{center}

\caption{Examples of \rbox{} measurements for six galaxies. In each panel, the
red arrows indicate the position angle and size of the boxy region, $2 \times
\rbox$; the longer cyan arrows indicate the position angle and full size of the bar.
Thin black dashed lines indicate position angle of line of nodes.
North is up and East is to the left in all panels. See also
Figures~\ref{fig:edge-on-with-sizes} and \ref{fig:A4B4-real-galaxies},  and the
Appendix. (See Appendix for image sources.)\label{fig:rbox-demos}}

\end{figure*}

\begin{figure*}
\begin{center}
\includegraphics[scale=0.89]{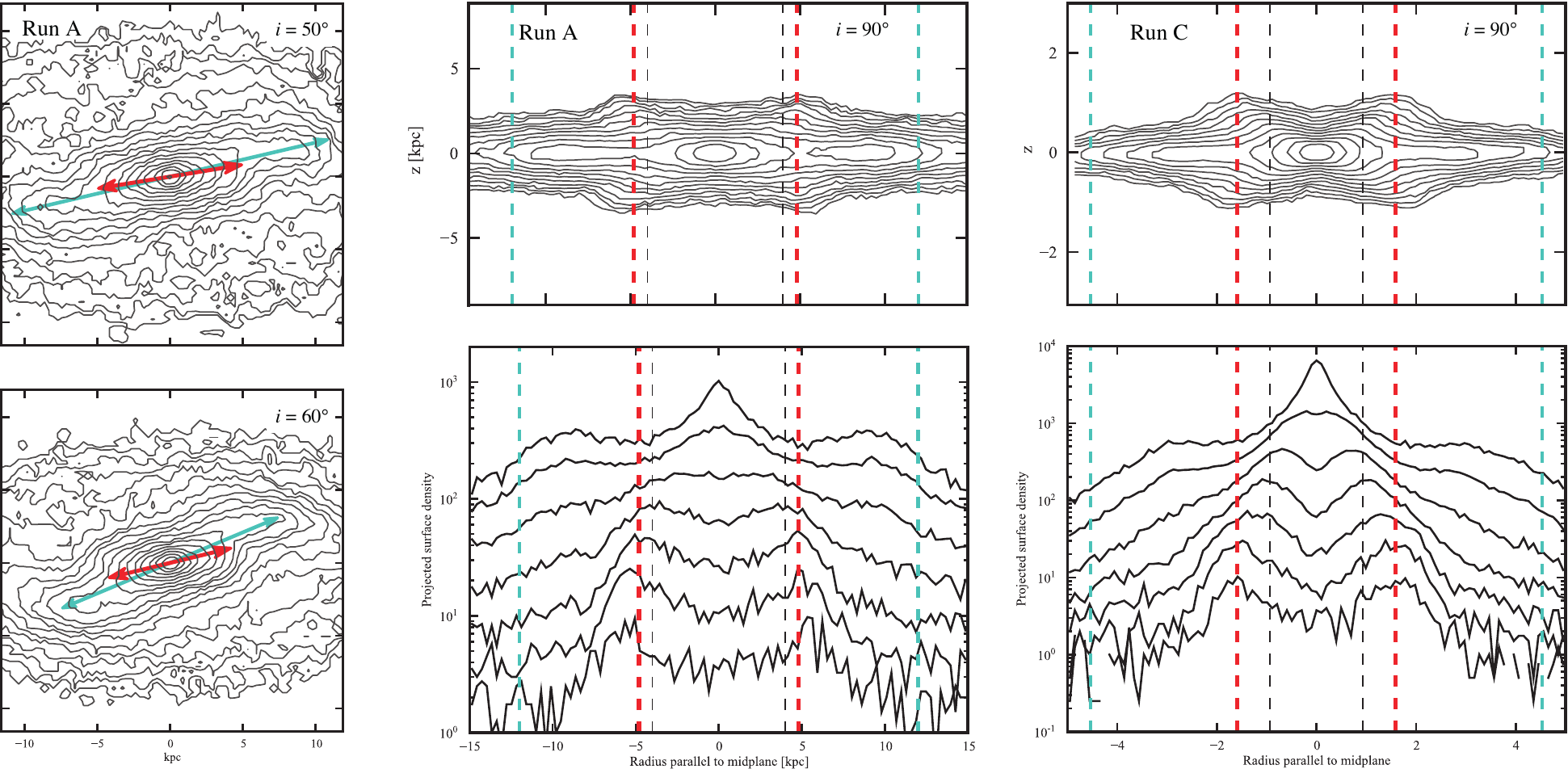}
\end{center}

\caption{Matching 2D morphology in moderate-inclination projections with 3D
morphology. \textbf{Left panels:} Projections of $N$-body model A, at $i =
50\degr$, bar $\dpa = 20\degr$ (top) and $i = 60\degr$, bar $\dpa = 40\degr$
(bottom). Red arrows mark the measured \rbox{} values; longer cyan arrows mark
the full bar radius. \textbf{Middle panel -- top:} Edge-on view of the same
simulation, with the bar oriented perpendicular to the line of sight.
\textbf{Middle panel -- bottom:} Cuts through the edge-on view, parallel to the
galaxy midplane (from top to bottom, the profiles are at $|z| = 0$, 0.9, 1.5,
2.1, 2.7, 3.3, and 3.9 kpc). In both panels, thin vertical black lines mark the
min($z_{4}$) radius, vertical red lines mark the average (deprojected) value of
\rbox{} from measurements on moderately inclined projections (e.g., left
panels), and vertical cyan lines mark full bar radius. \textbf{Right panels:}
Same as middle panels, but showing model C at $t = 600$ (parallel
cuts at $|z| = 0$, 0.3, 0.5, 0.7, 0.9, 1.1, and 1.3). Taken together, these show
that \rbox, measured on moderately inclined images, is a reasonable estimate of
the extent of the B/P structure, as seen in the edge-on
views.\label{fig:edge-on-with-sizes}}

\end{figure*}

After considerable experimentation, we settled on a direct visual measurement of
the extent of the boxy region: \rbox. This is the radius from the centre of the
galaxy along the bar major axis (more specifically, along the major axis of the
boxy-region isophotes) beyond which the spurs dominate. There is inevitably some
ambiguity in measuring this radius, but we find that it can usually be
determined with a precision of $\sim 10$\%, which is at least roughly comparable
to the uncertainty in determining the overall length of the bar. Examples of
\rbox{} measurements on real galaxies are given in Figures~\ref{fig:rbox-demos},
while examples using projected $N$-body models are given in
Figure~\ref{fig:edge-on-with-sizes}. Additional examples using the isophotes of
real galaxies are presented in Figure~\ref{fig:A4B4-real-galaxies} and in the
Appendix.

% wide figure -- use figure*

Figure~\ref{fig:edge-on-with-sizes} shows measurements of \rbox{} on two
moderately inclined projections of run A (left-hand panels). The middle panels
of that figure show something inaccessible for real galaxies with moderate
inclinations: the edge-on view of the simulation, with the bar perpendicular to
the line of sight, showing the full B/P structure. Parallel cuts through the
edge-on view are shown in the bottom middle panel. We measured \rbox{} on a
number of different projections (varying inclination and bar \dpa); the mean of
the deprojected values for this simulation was 4.8 kpc. This radius is marked in
the middle panels by the vertical dashed red lines. (The right-hand panels show
the result of the same exercise for another simulation.) Although one could
argue that the deprojected \rbox{} measurement slightly underestimates the full
radial extent of the B/P structure, it is none the less a surprisingly good
match. The very upper part of the B/P structure may extend slightly beyond the
boxy zone into the spurs, but the majority of the stars making up the bar at
this radius are in planar orbits, so the bar is predominantly flat at this
point.

In Figure~\ref{fig:edge-on-with-sizes} we also plot the radius where $z_{4}$,
the fourth-order Gauss-Hermite moment of the vertical density distribution along the
bar major axis, reaches a minimum. This is a measurement of the B/P structure
used by \citet{debattista05}\footnote{Referred to as $d_{4}$ in that paper.} and
\citet{mendez-abreu08}, who found that it closely matched the minimum in
$h_{4}$, the fourth-order Gauss-Hermite moment of the stellar \textit{velocity}
distribution, something which could be measured in face-on bars using
spectroscopy. We note that the radius of minimum $z_{4}$ is usually
\textit{smaller} than \rbox, something which should be kept in mind when
comparing spectroscopic measurements of face-on bars with our morphological
measurements.

% Final version -- uses edge-on_isophotes_and_cuts_alt2.ai

\subsection{Can We Use Ellipse Fits to Identify and Measure Inclined Box/Peanut Structures?}\label{sec:efits}

Strong versions of the boxy-bar morphology (e.g., Figures~\ref{fig:box+spurs}
and \ref{fig:offset-spurs}) are rather easy to spot from visual inspection of
the isophotes, and measuring the size of the boxy region on  images of such
galaxies is not too difficult.  It would clearly be desirable, however, to have
a consistent set of criteria which could be applied in a semi-automated fashion
to images, so that one could more easily identify weaker examples. Since the
process of fitting ellipses to galaxy isophotes is widespread and easily done,
it would be convenient if we could use ellipse fits for this purpose, and even
more so if we could define a way to measure \rbox{} using ellipse fits. The fact
that ellipse fits have traditionally been used to identify ``boxy'' isophotes in
elliptical galaxies would seem to suggest that they could be useful here as
well.

Unfortunately, considerable experimentation with ellipse fits to
isophotes of both projected $N$-body models and real galaxies has forced
us to conclude that ellipse-fitting does not provide a simple solution.
While we \textit{can} devise a set of criteria which will often -- but
not always -- indicate the \textit{presence} of a boxy bar (and thus the
projected B/P structure), attempts to devise a simple measurement of the
boxy zone's size run into problems.\footnote{This discussion is based on
ellipse fits of galaxies listed in Table~\ref{tab:galaxies}, galaxies
found in the analysis of our ``local sample'' in
Section~\ref{sec:local-sample}, and $N$-body models.}

We digress briefly to remind the reader of how ellipse fits are constructed and
analysed. The process of ellipse fitting involves finding an ellipse of a given
semi-major axis $a$ which best fits a given galaxy isophote, given the ability
to vary the ellipse's centre, position angle, and semi-minor axis $b$. The
particular implementation we use is that of the \textsc{iraf} task
\textsc{ellipse}, part of the \textsc{stsdas} package and based on the approach
of \citet{jz87}.   If an ellipse is a perfect fit to the isophote, then the
intensity along the ellipse will be constant. In practice, this is never true,
so the variations in intensity along the ellipse can be expanded as a Fourier
sum:
\begin{equation}
I(\theta) = I_{0} + \sum_{n = 1}^{\infty} [\tilde{A}_{n} 
\sin n\theta + \tilde{B}_{n} \cos n\theta],
\end{equation}
where $\theta$ is the eccentric anomaly. For a best-fitting ellipse, the first-
and second-order coefficients will be zero.  In order to describe how the
isophote differs \textit{spatially} from the fitted ellipse, the higher-order
($n \geq 3$) coefficients are divided by the local radial intensity gradient and
by the ellipse semi-major axis.  This transforms them into normalized
coefficients of \textit{radial} deviation $\delta r$ from a perfect ellipse, in
a coordinate system where the fitted ellipse is a circle with radius $r =
(ab)^{1/2}$:
\begin{equation}
\frac{\delta r(\theta)}{r} = \sum_{k = 3}^{4} [A_{k} \sin k
\theta + B_{k} \cos k \theta].
\end{equation}

The most commonly used higher-order coefficient is \bfour, the
$\cos 4 \theta$ term, which measures symmetric distortions from pure ellipticity
along the ellipse major axis.  When $B_{4} > 0$, the isophotes are pointed or
``discy''; when $B_{4} < 0$, the isophotes have a more rectangular or ``boxy''
shape. Note that some other ellipse-fitting codes (e.g., that of \nocite{bender88}Bender,
D\"{o}bereiener, \& M\"{o}llenhoff 1988), designate the $\sin 4 \theta$ and
$\cos 4 \theta$ terms by $b_{4}/a$ and $a_{4}/a$, respectively.\footnote{The
conversion between the different $\cos 4 \theta$ coefficients is $a_{4}/a =
\sqrt{b/a} \, B_{4}$ \citep{bender88}.}

\begin{figure*}
\begin{center}
\includegraphics[scale=0.8]{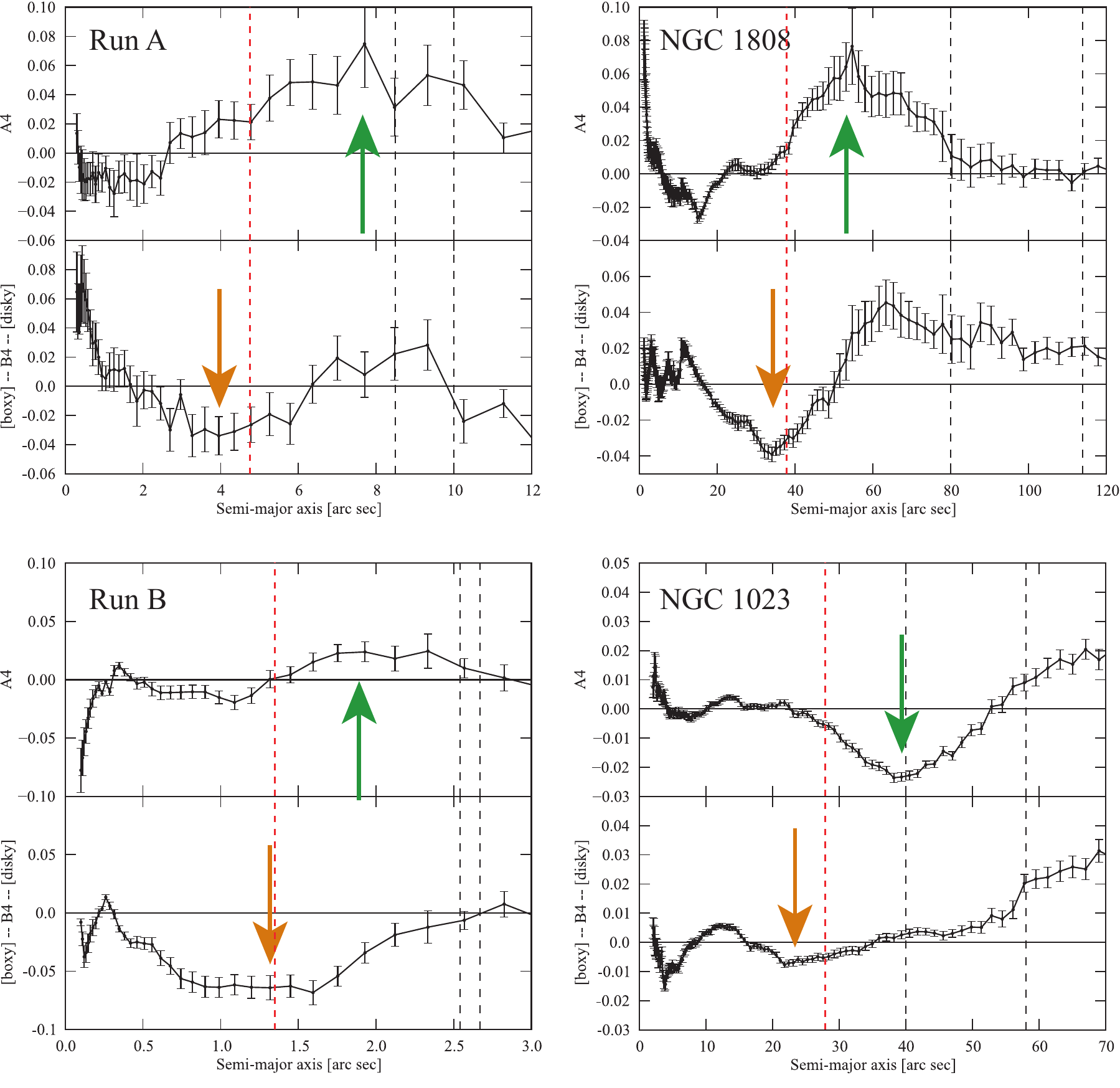}
\end{center}

\caption{Signatures of box+spurs morphology in ellipse fits, using $\afour$
($\sin 4 \theta$) and $\bfour$ ($\cos 4 \theta$) coefficients. Vertical dashed
black lines mark semi-major axes of our two estimates of bar size (\amax{} and \lbar);
vertical short-dashed red lines mark \rbox, our visual measurement of the size of the boxy region.
\textbf{Left panels:} Ellipse fits to $N$-body simulations. Upper panel: run A,
seen with $i = 60\degr$, $\dpa = 20\degr$; lower panel: run B at $t = 1000$, seen
with $i = 45\degr$, $\dpa = 40\degr$. \textbf{Right panels:} Ellipse fits to
real galaxies. Upper panel: NGC~1808, a strong boxy-bar case (see
Figure~\ref{fig:rotating-spurs}); lower panel: NGC~1023, a much weaker case. In
all panels, there a pattern of boxy isophotes ($\bfour < 0$) at small radii
(gold arrows), transitioning at larger radii to discy isophotes ($\bfour > 0$)
as the spurs become more prominent; at the same time, \afour{} becomes strongly
nonzero closer to the bar end (green arrows), indicating the offset orientation
of the spurs. \label{fig:A4B4-1}}

\end{figure*}

One might expect that the boxy zone would be marked in the ellipse fits by
negative \bfour{} values, transitioning to more elliptical -- even discy --
isophotes outside. This is often true, and Figure~\ref{fig:A4B4-1}
shows some examples of the pattern. In some galaxies, however, the isophotes may
never become boxy enough to acquire negative \bfour{} values.

\begin{figure*}
\begin{center}
\includegraphics[scale=0.8]{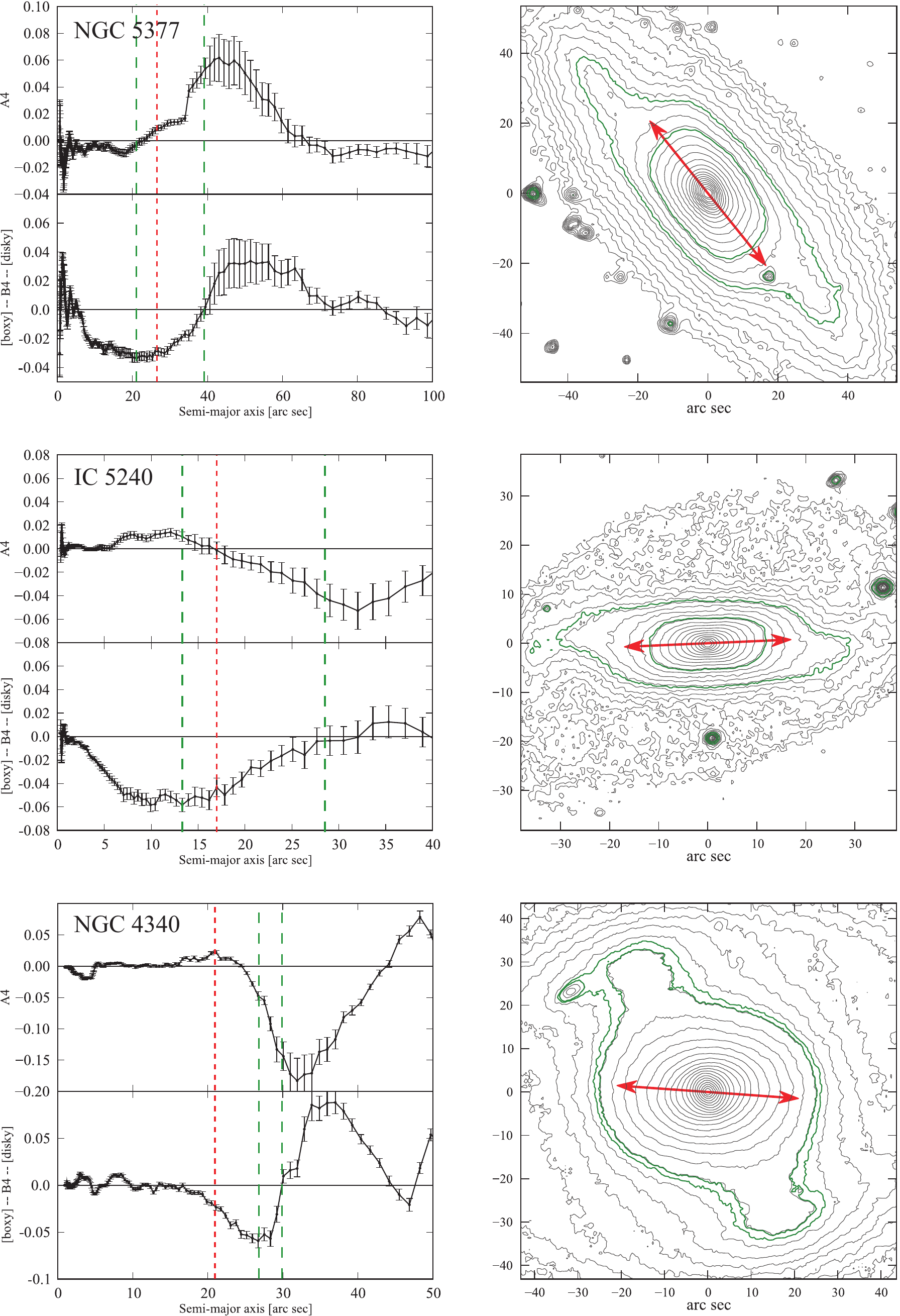}
\end{center}

\caption{Why ellipse fits are problematic for measuring B/P sizes.
\textbf{Left}: $\afour$ ($\sin 4 \theta$) and $\bfour$ ($\cos 4 \theta$)
coefficients from ellipse fits. Vertical green long-dashed lines mark semi-major
axes corresponding to the maximally boxy (min$(\bfour)$) isophote and the
isophote where $\bfour = 0$ immediately outside. Vertical short-dashed red lines
indicate \rbox, our visual measurement of the size of the boxy region.
\textbf{Right}: Log-scaled isophotes of NGC 5377 (IRAC1, outer disc excluded),
IC~5240 ($K$-band), and NGC~4340 (SDSS $r$-band); N is up and E is to the left.
Red arrows mark the boxy region ($2 \times \rbox$); thicker green lines  outline
isophotes corresponding to the min$(\bfour)$ and $\bfour = 0$ fitted ellipses.
Neither min$(\bfour)$ nor $\bfour = 0$ can be used to reliably and accurately
define the limits of the boxy region.\label{fig:A4B4-real-galaxies}}

\end{figure*}

The \afour{} term (the $\sin 4 \theta$ coefficient) is also useful, because it
can indicate the presence of offset spurs.  Nonzero $A_{4}$ values mean that the
fitted isophote has deviations from bisymmetry (e.g., transforming the symmetric
rectangular shape into something more like a parallelogram): when $A_{4} > 0$,
the isophote ends are offset counter-clockwise from the major
axis of the fitted ellipse, and when $A_{4} < 0$, they are offset clockwise; see
Figure~\ref{fig:A4B4-real-galaxies}.

Thus, a reasonable set of criteria for identifying the boxy-bar morphology might
include the following:
\begin{enumerate}
  \item The presence of an inner boxy region: $\bfour < 0$ somewhere 
  inside the bar. (In some cases, the ``boxy'' region will be close to
  elliptical, with $\bfour \approx 0$.)
  
  \item This region  corresponds to a
  value of \afour{} near zero and (sometimes) to a plateau or shoulder in the 
  position-angle profile.  This is the region of symmetric, boxy isophotes, 
  corresponding to the projected box/peanut.
  
  \item At larger radii (but still inside the bar), the  isophotes become
  \textit{discy} ($\bfour > 0$); this is the region of the spurs outside the
  boxy zone, corresponding to the flat part of the bar outside the box/peanut.
  
  \item Almost always, the \afour{} term becomes significantly nonzero in the same
  region, and the position angle continues to change; in at least some cases, the 
  extremum in \afour{} happens slightly \textit{inside} the peak in \bfour.
  This is the signature of \textit{offset} spurs, which indicates that the bar is
  not aligned with the galaxy line of nodes. (Spurs which are aligned
  would be indicated by $\afour  = 0$; this means that the bar lies along
  the line of nodes.)
\end{enumerate}

The preceding set of criteria suggest an appealingly simple correspondence: boxy
zone = boxy isophotes (i.e., $\bfour < 0$), spurs = discy isophotes (i.e.,
$\bfour > 0$). So could we simply use the semi-major axis of maximum boxyness
(minimum \bfour) to derive \rbox{}? Or, alternately, could we use the semi-major
axis where \bfour{} crosses from negative to positive?

In practice, this simple idea does not work for most galaxies. Figure~\ref{fig:A4B4-real-galaxies}
shows that the min(\bfour) measurement usually \textit{underestimates} \rbox{}
(the size of the boxy zone).  And the $\bfour = 0$ semi-major axis turns
out to correspond to an isophote which is actually well into the spur-dominated
region, thus strongly \textit{overestimating} \rbox. In other words, even
outside the boxy region, where the spurs are clearly present, the best-fitting
ellipses can be sufficiently affected by the boxy region so as to have boxy
deviations.\footnote{We also observe that it is in general not wise to assume that
isophotes with $\bfour = 0$ are always actually \textit{elliptical}; they can be
strongly non-elliptical.}

As a compromise, we have found that the \textit{mean} of $a({\rm min}(\bfour))$
and $a(\bfour = 0)$  is often a reasonable approximation of \rbox. So if one
\textit{must} use an ellipse-fit-based method for estimating \rbox, one could
certainly do worse than to use this. It \textit{does} start to fail
systematically when the bar is close to the galaxy minor axis, however
(overestimating \rbox; see NGC~4340 in Figure~\ref{fig:A4B4-real-galaxies}), so
we recommend measuring \rbox{} directly on the image or isophotes whenever
possible.

\section{Exploring a Local Sample}\label{sec:local-sample}

While the list of boxy-bar galaxies in Table~\ref{tab:galaxies} is of potential
use in providing candidates for detailed individual investigations, its
heterogenous nature tells us little about how common such features actually are.
In this section, we focus on a well-defined local sample of barred galaxies and
attempt to determine the frequency of the boxy-bar phenomenon.

The sample we use consists of nearby S0--Sb galaxies with bars, taken from
the combined S0--Sb sample presented in \citet{erwin08} and \citet{gutierrez11}.
This sample was defined so as to include all galaxies from the UGC catalog
\citep{ugc} which met the following criteria: RC3 major-axis diameter $D_{25}
\geq 2.0\arcmin$, RC3 axis ratio $a/b \leq 2.0$, redshift $V \leq 2000$ \kms{}
(from NED), and declination $\geq -10\degr$; S0 galaxies in the Virgo Cluster
were also included \citep[based on membership in the Virgo Cluster
Catalog;][]{binggeli85}, ignoring the redshift limit. This produced a total of 122
galaxies, of which nine were excluded for being highly disturbed (e.g., merger
remnants or polar-ring systems) or edge-on despite their low axis ratios (e.g.,
S0 galaxies with large bulges); see \citet{erwin05} and \citet{gutierrez11} for
specifics.  Of the remaining 113 galaxies, 78 proved to have bars; measurements
of the bar parameters (size, position angle, maximum isophotal ellipticity) are
presented in \citet{erwin05}, \citet{erwin08}, and \citet{gutierrez11}, along
with disc measurements.\footnote{For NGC~2712, we use an updated disc position
angle of 178\degr.}

An axis ratio limit of $a/b \leq 2.0$ is the same as that commonly used to
maximize the identification (and measurement) of bars, and formally
corresponds to inclinations $\la 62\degr$, assuming an intrinsic axis ratio of
$c/a = 0.2$.  Detailed analyses of individual galaxies showed that some probably
have inclinations as high at 66\degr, but we did not attempt to exclude these
systems.

One of the main virtues of such a locally defined sample is the high spatial
resolution it affords. Even when we are restricted to images with seeing FWHM
$\sim 1.5$--2\arcsec, this is significantly smaller than the typical sizes of
bars in our sample \citep[see][]{erwin05,gutierrez11}; in addition, \textit{HST}
images are available for many of the galaxies, which helps with resolving the
structure of the smaller bars.

% This table is generated by
% $ ./make_tables.py --wiyn

\begin{table*}
\begin{minipage}{126mm}
    \caption{Local Barred-Galaxy Sample}
    \label{tab:wiyn}
    \begin{tabular}{llrlrrllll}
\hline
Name & Type & Distance & $M_{B}$  & $i$  & \dpa  & Boxy    & Spurs & Offset & Lead/Trail \\
(1)  & (2)  & (3)      & (4)      & (5)  & (6)     & (7)     & (8)   & (9)    & (10)      \\
\hline
 NGC 278 &                 SAB(rs)b &  11.0 & $-19.40$ &   17 &   51 &   no &   no &  --- &  --- \\ 
 NGC 718 &                  SAB(s)a &  22.6 & $-19.43$ &   30 &   37 &   no &   no &  --- &  --- \\ 
 NGC 936 &            SB(rs)$0^{+}$ &  23.0 & $-20.86$ &   41 &   57 &   no &   no &  --- &  --- \\ 
NGC 1022 &     (R$^{\prime}$)SB(s)a &  18.1 & $-19.46$ &   24 &   61 &  yes &  yes &  yes & trail \\ 
NGC 1068 &               (R)SA(rs)b &  14.2 & $-21.23$ &   31 &   76 &   no &   no &  --- &  --- \\ 
NGC 2273 &                  SB(r)a: &  27.3 & $-20.11$ &   50 &   74 &   no &   no &  --- &  --- \\ 
NGC 2681 & (R$^{\prime}$)SAB(rs)0/a &  17.2 & $-20.20$ &   18 &   71 &   no &   no &  --- &  --- \\ 
NGC 2712 &                  SB(r)b: &  26.5 & $-19.88$ &   59 &   53 &  yes & yes? &  yes & trail \\ 
NGC 2787 &             SB(r)$0^{+}$ &   7.5 & $-18.20$ &   55 &   65 &   no &   no &  --- &  --- \\ 
NGC 2859 &          (R)SB(r)$0^{+}$ &  24.2 & $-20.21$ &   32 &   79 &   no &   no &  --- &  --- \\ 
NGC 2880 &                SB$0^{-}$ &  21.9 & $-19.38$ &   52 &   72 &   no &   no &  --- &  --- \\ 
NGC 2950 &          (R)SB(r)$0^{0}$ &  14.9 & $-19.14$ &   48 &   48 &  no? &  no? &  --- &  --- \\ 
NGC 2962 &        (R)SAB(rs)$0^{+}$ &  30.0 & $-19.71$ &   53 &   30 & yes? & yes? &  yes &  --- \\ 
NGC 3031 &                  SA(s)ab &   3.6 & $-20.73$ &   58 &   18 &  yes & yes? &  yes & lead \\ 
NGC 3049 &                 SB(rs)ab &  20.2 & $-18.65$ &   51 &    8 &   no &   no &  --- &  --- \\ 
NGC 3185 &                (R)SB(r)a &  17.5 & $-18.61$ &   49 &   37 &  yes &  yes &  yes & lead \\ 
NGC 3351 &                   SB(r)b &  10.0 & $-19.94$ &   46 &   82 &   no &   no &  --- &  --- \\ 
NGC 3368 &                SAB(rs)ab &  10.5 & $-20.37$ &   50 &   67 &  yes & yes? & yes? & lead? \\ 
NGC 3412 &             SB(s)$0^{0}$ &  11.3 & $-18.98$ &   58 &   68 &   no &   no &  --- &  --- \\ 
NGC 3485 &                  SB(r)b: &  20.0 & $-19.03$ &   26 &   43 &   no &  no? &  --- &  --- \\ 
NGC 3489 &           SAB(rs)$0^{+}$ &  12.1 & $-19.45$ &   58 &   72 &   no &   no &  --- &  --- \\ 
NGC 3504 &              (R)SAB(s)ab &  22.3 & $-20.29$ &   22 &    6 &   no &   no &  --- &  --- \\ 
NGC 3507 &                   SB(s)b &  14.2 & $-19.21$ &   27 &   24 &   no &   no &  --- &  --- \\ 
NGC 3599 &                SA$0^{0}$ &  19.8 & $-18.70$ &   22 &   62 &   no &   no &  --- &  --- \\ 
NGC 3626 &         (R)SA(rs)$0^{+}$ &  19.5 & $-19.75$ &   49 &   11 & yes? &  yes &  yes & lead \\ 
NGC 3729 &               SB(r)a pec &  16.8 & $-19.35$ &   53 &   50 &  yes &  yes &  yes & lead \\ 
NGC 3941 &             SB(s)$0^{0}$ &  12.2 & $-19.31$ &   52 &   33 & yes? & yes? &  yes &  --- \\ 
NGC 3945 &         (R)SB(rs)$0^{+}$ &  19.8 & $-19.94$ &   55 &   88 &   no &   no &  --- &  --- \\ 
NGC 3982 &                 SAB(r)b: &  20.9 & $-19.95$ &   29 &    8 &   no &   no &  --- &  --- \\ 
NGC 3998 &             SA(r)$0^{0}$ &  13.7 & $-19.36$ &   38 &   13 &   no &   no &  --- &  --- \\ 
NGC 4037 &                 SB(rs)b: &  13.5 & $-17.79$ &   32 &   46 &  yes & yes? & yes? & trail \\ 
NGC 4045 &                  SAB(r)a &  26.8 & $-19.70$ &   48 &   78 &  no? &  no? &  --- &  --- \\ 
NGC 4102 &                 SAB(s)b? &  14.4 & $-19.22$ &   55 &   44 & yes? & yes? &  yes & trail \\ 
NGC 4143 &            SAB(s)$0^{0}$ &  15.9 & $-19.40$ &   59 &   34 & yes? &  no? &  --- &  --- \\ 
NGC 4151 & (R$^{\prime}$)SAB(rs)ab: &  15.9 & $-20.70$ &   20 &   73 &   no &   no &  --- &  --- \\ 
NGC 4203 &               SAB$0^{-}$ &  15.1 & $-19.21$ &   28 &    2 &   no &   no &  --- &  --- \\ 
NGC 4245 &                 SB(r)0/a &  12.0 & $-18.28$ &   38 &   43 &   no &  no? &  --- &  --- \\ 
NGC 4267 &            SB(s)$0^{-}$? &  15.3 & $-19.25$ &   25 &   86 &   no &   no &  --- &  --- \\ 
NGC 4314 &                  SB(rs)a &  12.0 & $-19.12$ &   25 &   82 &   no &   no &  --- &  --- \\ 
NGC 4319 &                  SB(r)ab &  23.5 & $-19.26$ &   42 &   22 &  yes & yes? & yes? & lead \\ 
NGC 4340 &             SB(r)$0^{+}$ &  15.3 & $-18.90$ &   50 &   73 &  yes &  yes &  yes &  --- \\ 
NGC 4369 &               (R)SA(rs)a &  16.6 & $-18.84$ &   18 &   78 &   no &   no &  --- &  --- \\ 
NGC 4371 &             SB(r)$0^{+}$ &  15.3 & $-19.32$ &   58 &   85 &   no &   no &  --- &  --- \\ 
NGC 4386 &              SAB$0^{0}$: &  27.0 & $-19.68$ &   48 &    9 & yes? & yes? &  yes &  --- \\ 
NGC 4477 &           SB(s)$0^{0}$:? &  15.3 & $-19.69$ &   33 &   71 &   no &  no? &  --- &  --- \\ 
NGC 4531 &               SB$0^{+}$: &  15.2 & $-18.67$ &   49 &   38 &   no &   no &  --- &  --- \\ 
NGC 4596 &             SB(r)$0^{+}$ &  15.3 & $-19.63$ &   42 &   55 &   no &   no &  --- &  --- \\ 
NGC 4608 &             SB(r)$0^{0}$ &  15.3 & $-19.02$ &   36 &   78 &   no &   no &  --- &  --- \\ 
NGC 4612 &            (R)SAB$0^{0}$ &  15.3 & $-19.01$ &   44 &   67 &   no &   no &  --- &  --- \\ 
NGC 4643 &                SB(rs)0/a &  18.3 & $-19.85$ &   38 &   82 &   no &   no &  --- &  --- \\ 
NGC 4665 &                 SB(s)0/a &  10.9 & $-18.87$ &   26 &   66 &   no &   no &  --- &  --- \\ 
NGC 4691 &          (R)SB(s)0/a pec &  15.1 & $-19.43$ &   38 &   58 &   no &   no &  --- &  --- \\ 
NGC 4699 &                 SAB(rs)b &  18.9 & $-21.37$ &   42 &   17 & yes? &  yes & yes? & trail? \\ 
NGC 4725 &             SAB(r)ab pec &  12.4 & $-20.69$ &   42 &   13 &  yes &  yes &  yes & trail \\ 
\hline
\end{tabular}

\medskip

Notes on the presence or absence of boxy-bar features in a local sample of
S0--Sb barred galaxies. (1) Galaxy name. (2) Hubble type from RC3. (3) Distance
in Mpc \citep[for sources, see][]{erwin08,gutierrez11}. (4) Absolute $B$
magnitude, from HyperLeda $B_{tc}$ and our adopted distance. (5) Galaxy inclination.
(6) Deprojected angle between bar and disc major axis. (7) Indicates whether
bar displays boxy interior. (8) Indicates whether narrow spurs outside boxy
interior are seen. (9) Indicates whether spurs, if present, are offset from
major axis of boxy interior. (10) Indicates whether offset spurs, if present,
lead or trail (assuming main spiral pattern is trailing).

\end{minipage}
\end{table*}

\setcounter{table}{1}

\begin{table*}
\begin{minipage}{126mm}
    \caption{Continued}
%    \label{tab:wiyn}
    \begin{tabular}{llrlrrllll}
\hline
Name & Type & Distance & $M_{B}$  & $i$  & \dpa  & Boxy    & Spurs & Offset & Lead/Trail \\
(1)  & (2)  & (3)      & (4)      & (5)  & (6)     & (7)     & (8)   & (9)    & (10)      \\
\hline
NGC 4736 &               (R)SA(r)ab &   5.1 & $-19.98$ &   35 &   27 &   no &   no &  --- &  --- \\ 
NGC 4750 &              (R)SA(rs)ab &  25.4 & $-20.27$ &   30 &   43 &   no &   no &  --- &  --- \\ 
NGC 4754 &            SB(r)$0^{-}$: &  16.8 & $-19.78$ &   62 &   75 &   no &   no &  --- &  --- \\ 
NGC 4772 &                   SA(s)a &  14.5 & $-19.22$ &   44 &   16 &  no? &  no? &  --- &  --- \\ 
NGC 4941 &             (R)SAB(r)ab: &  15.0 & $-19.37$ &   48 &    6 &  no? &  no? &  --- &  --- \\ 
NGC 4995 &                  SAB(r)b &  23.6 & $-20.41$ &   47 &   74 & yes? &  yes &  yes & trail \\ 
NGC 5338 &               SB$0^{0}$: &  12.8 & $-16.70$ &   66 &   55 &  no? &  no? &  --- &  --- \\ 
NGC 5377 &                (R)SB(s)a &  27.1 & $-20.29$ &   59 &   35 &  yes &  yes &  yes & trail \\ 
NGC 5701 &             (R)SB(rs)0/a &  21.3 & $-19.97$ &   20 &   50 &   no &   no &  --- &  --- \\ 
NGC 5740 &                 SAB(rs)b &  22.0 & $-19.67$ &   60 &   57 &  yes &  yes &  yes & trail \\ 
NGC 5750 &                 SB(r)0/a &  26.6 & $-19.94$ &   62 &   72 &  yes & yes? & yes? & trail \\ 
NGC 5806 &                  SAB(s)b &  19.2 & $-19.67$ &   58 &   13 &  yes &  yes &  yes & trail \\ 
NGC 5832 &                 SB(rs)b? &   9.9 & $-17.15$ &   55 &   76 &  no? &  no? &  --- &  --- \\ 
NGC 5957 &    (R$^{\prime}$)SAB(r)b &  26.2 & $-19.36$ &   15 &    3 &  no? &  no? &  --- &  --- \\ 
NGC 6012 &              (R)SB(r)ab: &  26.7 & $-19.78$ &   33 &   59 &   no &   no &  --- &  --- \\ 
NGC 6654 &   (R$^{\prime}$)SB(s)0/a &  28.3 & $-19.65$ &   44 &   23 &   no &  no? &  --- &  --- \\ 
NGC 7177 &                  SAB(r)b &  16.8 & $-19.79$ &   48 &   76 &  no? &  no? &  --- &  --- \\ 
NGC 7280 &            SAB(r)$0^{+}$ &  24.3 & $-19.16$ &   50 &   28 &   no &   no &  --- &  --- \\ 
NGC 7743 &          (R)SB(s)$0^{+}$ &  20.7 & $-19.49$ &   28 &   11 & yes? & yes? &  yes & lead \\ 
  IC 499 &                       Sa &  29.5 & $-19.37$ &   59 &   47 &  no? &   no &  --- &  --- \\ 
  IC 676 &          (R)SB(r)$0^{+}$ &  19.4 & $-18.42$ &   47 &   41 &  no? & yes? & yes? & lead \\ 
 IC 1067 &                   SB(s)b &  22.2 & $-18.82$ &   44 &   40 &  yes &  yes &  yes & trail \\ 
UGC 3685 &                  SB(rs)b &  26.8 & $-19.51$ &   31 &   14 &   no &   no &  --- &  --- \\ 
UGC 11920 &                    SB0/a &  18.0 & $-19.71$ &   52 &    8 &  no? &  no? &  --- &  --- \\ 
\hline
\end{tabular}

\end{minipage}
\end{table*}

\subsection{Analysis}

We analysed the best available images for all galaxies in the sample to
determine if they showed evidence for the boxy-bar morphology; we counted both
the strong examples discussed in Section~\ref{sec:boxy-bars} and weaker examples
suggested by some of the $N$-body projections (e.g., cases where the spurs are
short and/or project from corners of the boxy zone). Our primary method of
analysis was visual inspection of the images, and of isophote contour plots
derived from the images. (The suggested ellipse-fit-based method we discuss in
Section~\ref{sec:efits} was derived \textit{after} this analysis, using galaxies
identified visually, including those found in this sample.)

For dust-free S0 galaxies, we generally used red ($R$, $r$, or $i$) optical
images from the SDSS (DR7) or from other sources discussed in \citet{erwin03},
\citet{erwin08}, and \citet{gutierrez11}. For galaxies with dust obscuration in
the bar region -- including almost all of the spiral galaxies -- we used near-IR
imaging from a variety of sources, the most common being \textit{Spitzer} IRAC1
($3.6\mu$m) images from NED or from the \textit{Spitzer} archive. Most of the
\textit{Spitzer} images are part of the \textit{Spitzer} Survey of Stellar
Structure in Galaxies \citep[S$^{4}$G;]{sheth10}; other sources included SINGS
\citep{kennicutt03}, and the \textit{Spitzer} Local Volume Legacy
\citep{dale09}. We also used $H$ and $K$ images available from NED (mostly
higher in resolution or S/N than IRAC1 images), including those from
\citet{knapen03}, \citet{eskridge02}, \citet{mh01}, and \citet{wu02}, and a set of $J$ and $H$
images taken with the INGRID imager on the William Herschel Telescope
\citep[e.g.,][]{erwin03-id,nowak10}. Finally, for some galaxies with
particularly small bars (e.g., NGC 4102) we used archival \textit{HST} NICMOS2
and NICMOS3 images, mostly obtained with the F160W filter. The median resolution of
the images we used was FWHM $= 1.1\arcsec$, with a range of 0.5--2.0\arcsec{}
(excluding the five galaxies for which we used \textit{HST} NICMOS2 or NICMOS3
images).

The primary results of our analysis are coded in Table~\ref{tab:wiyn}, where we
indicate whether or not the bar of each galaxy\footnote{In the case of
double-barred galaxies, we analyse the outer bar.} has a boxy interior, and if
so, whether it has spurs and whether the spurs are offset relative to the boxy
zone.  Less certain classifications are indicated by question marks. The offset
of the spurs is defined as leading or trailing based on the sense of spiral arm
rotation; this is not possible for some galaxies, such as S0 galaxies where the
absence of dust lanes and spiral arms prevents us from determining a sense of
rotation for the galaxy.

\begin{figure*}
\begin{center}
\includegraphics[scale=0.8]{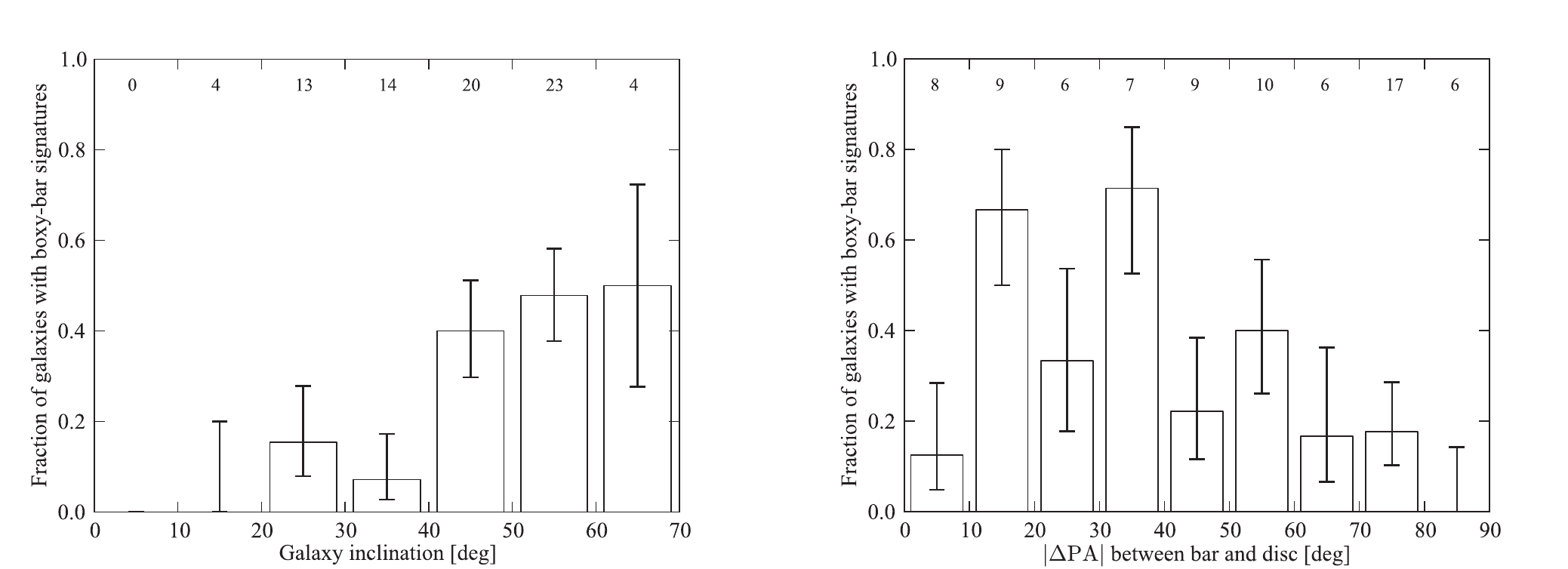}
\end{center}

\caption{\textbf{Left panel}: Fraction of galaxies in our local sample
(Table~\ref{tab:wiyn}) with detected boxy-bar morphologies, as a
function of galaxy inclination; the total number of galaxies in each
inclination bin is listed along the top of the figure. \textbf{Right
panel}: Same, but now showing fraction as a function of \dpa, the
(deprojected) relative angle between the bar and the galaxy major axis.
Error bars in both panels are binomial uncertainties calculated using
the method of \citet{wilson27}.\label{fig:wiyn-inc-deltapa}}

\end{figure*}

The left panel of Figure~\ref{fig:wiyn-inc-deltapa} shows how
the fraction of galaxies with boxy-bar morphology depends on galaxy
inclination.  As we would expect, the fraction rises as we go to higher
inclinations; for $i > 40\degr$, roughly half of the bars have boxy
interiors.  What is perhaps unexpected is how \textit{low} in
inclination one can go and still detect boxy interiors: there are two
galaxies with inclinations $20\degr < i < 30\degr$ where we find
evidence for the boxy-bar structure: NGC~1022 ($i = 24\degr$) and
NGC~7743 ($i = 28\degr$).

The right panel of the same figure shows the fraction of galaxies
which have boxy-bar morphologies as a function of \dpa{}, using the
deprojected position angles.   The boxy-bar morphology is clearly most
common when the relative position angle is between 10\degr{} and
40\degr{}, and is rare for $\dpa > 60\degr$.  If we divide the sample
into galaxies with $\dpa < 45\degr$ (35 total) and galaxies with $\dpa >
45\degr$ (43 total), the boxy-bar fractions are $43 \pm 8$\% and
$21^{+7}_{-6}$\%, respectively, though the statistical significance of
this difference is marginal (Fisher Exact Test $P = 0.049$).

% figure generated by
% $ ./histogram_figs_and_stats.py --inc
% $ ./histogram_figs_and_stats.py --barpa
% plus Illustrator (wiyn_boxybars_inc+dpa_2panel.ai)

\subsection{What Fraction of Candidate Galaxies Have B/P Bulges?}\label{sec:fraction}

Our analysis of the $N$-body simulations suggests that detection of the B/P
structure is maximized both when the inclination is high and when the bar is
closer to the major axis than to the minor axis. So if we are interested in
finding out how common B/P structures are in our local sample, it makes sense to
restrict ourselves to a subset of galaxies with reasonably high
inclinations and low values of \dpa{}.

For galaxies with $i \geq 40\degr$ and $\dpa < 45\degr$, we find that
$64^{+10}_{-11}$\% show at least a boxy interior; $59 \pm 10$\% of the subsample
show both a boxy interior and distinct spurs. (If we increase the inclination
limit to 50\degr, the frequencies become $70^{+12}_{-16}$\% and
$60^{+14}_{-16}$\%, respectively.) This suggests that, roughly speaking, at
least two-thirds of S0--Sb bars have buckled or otherwise thickened and produced
B/P structures. The fraction may well be higher if some of the bars have
relatively weak B/P structures, which do not produce a strong projected
signature when the inclination is lower; significant bulges or central discs can
potentially also weaken the apparent signature. We consider the possibility of
identifying individual galaxies which might lack B/P structures altogether in
Section~\ref{sec:not-buckled}.

\section{The Sizes of Box/Peanut Structures Relative to Bars}\label{sec:sizes}

For the 24 galaxies in our local sample where we found boxy-bar signatures, we
measured the size of the boxy region \rbox, as described in
Section~\ref{sec:3d}.   In absolute terms, \rbox{} ranges from 0.37 to 3.79 kpc,
with a mean of $1.46 \pm 0.91$ kpc; these sizes are deprojected using the
ellipse-fit position angle corresponding to the boxy region.

However, what is probably more interesting is the question of how much of any
given bar is vertically thickened. To investigate this, we calculated the size
of the boxy region relative to the length of the bar. All the local-sample
galaxies have bar measurements in \citet{erwin05} or \citet{gutierrez11}:
\amax{} and \lbar{} \citep[see][for definitions and comparisons with $N$-body
measurements]{erwin05}. Of the two bar-length measurements, \lbar{} is probably
more relevant, since it attempts to measure the full length of the bar; \amax{}
(the semi-major axis of maximum ellipticity) is a lower limit which in most
cases underestimates the true bar length. To compare \rbox{} with \lbar, we
deprojected both measurements; the deprojection of \rbox{} was as described in
the previous paragraph, while the deprojection of \lbar{} used the bar position
angle from \citet{erwin05} or \citet{gutierrez11}.

\begin{figure}
\begin{center}
\includegraphics[scale=0.44]{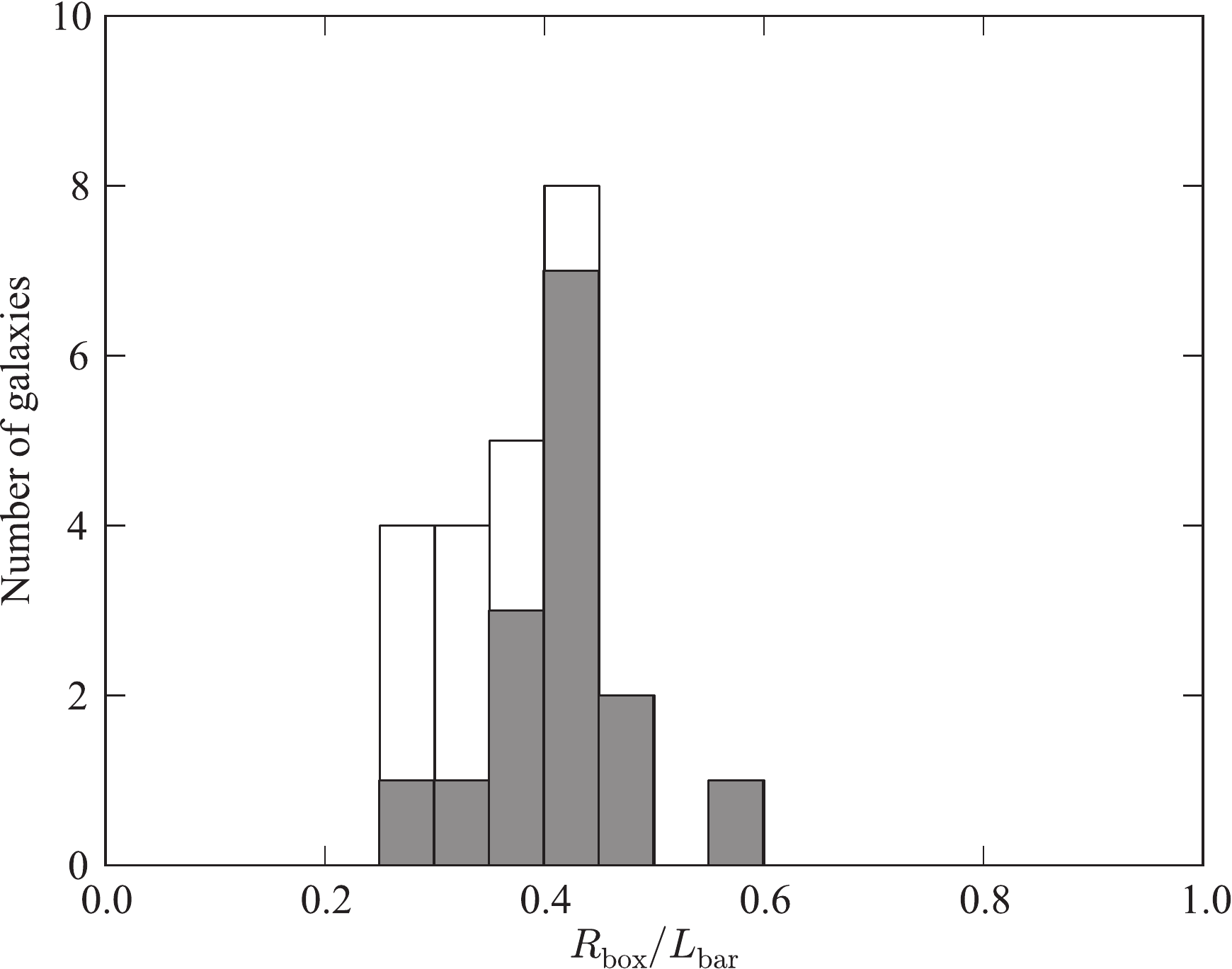}
\end{center}

\caption{Distribution of relative sizes (radius of B/P structure \rbox{} relative to
bar radius \lbar) for the local sample. Open bars: all 24 galaxies with detected
boxy-bar signatures. Grey bars: restricted to the 17 galaxies with (deprojected)
angle between bar and disc major axis $\leq 45\degr$, which maximizes detection of the
B/P structure.
\label{fig:relative-sizes-hist}}

\end{figure}

Figure~\ref{fig:relative-sizes-hist} shows the distribution of $\rbox/\lbar$ for
our local sample. For the complete sample, we find a mean size of $\rbox/\lbar =
0.38 \pm 0.08$ (median = 0.37); for the restricted subset of 15 galaxies with
$\dpa < 45\degr$, $\left<  \rbox/\lbar \right>  \, = \, 0.42 \pm 0.07$ (median =
0.43). Thus, it appears that on average box/peanut structures extend to slightly
less than half the bar length. Values of $\rbox/\lbar$ range from a low of 0.26
(0.29 for the $\dpa < 45\degr$ subset) to a maximum of 0.58. Given the
relatively narrow distribution in Figure~\ref{fig:relative-sizes-hist}, it is
not surprising that \rbox{} and \lbar{} are strongly correlated (Spearman $r =
0.92$, $P = 1.3 \times 10^{-10}$ for the complete local sample\footnote{$P$ =
probability of an $r$ value this high or higher  under the null hypothesis of no
true correlation.}). If we include an additional twelve galaxies not in our
sample where we have measured both \rbox{} and \lbar, the statistics do not
change significantly: $\left< \rbox/\lbar \right> \, = \, 0.38 \pm 0.07$ (median
= 0.38). If, instead, we use \amax{} for the bar size, the mean values are
$\rbox/\amax = 0.53 \pm 0.09$ for the local sample with $\dpa < 45\degr$ and
$\rbox/\amax = 0.47 \pm 0.11$ for all values of \dpa.

The only previous attempt to compare B/P sizes to bar sizes for a sample of
galaxies is that of \citet{lutticke00b}, who measured various structures in
near-IR images of edge-on galaxies. For six galaxies where there was a clear
peanut-shaped bulge and a shelf in the mid-plane surface-brightness profile
(suggesting a bar), they measured both ``BPL'' (the box/peanut length) and
``BAL'' (the bar length). Since the latter was measured at the point where the
bar excess appeared to merge with the outer exponential disc, it probably
corresponds (approximately) to our $\lbar$. Inverting their BAL/BPL measurements
to get an equivalent to $\rbox/\lbar$ yields a median value of 0.38 and a mean
of $0.38 \pm 0.06$. This is essentially identical to our findings when we use
\lbar{} to define the bar size, and is a nice confirmation of the idea that our
measurement of \rbox{} in moderately inclined galaxies does indeed map to
measurements of the off-plane structures of edge-on galaxies.

The mean value and range of relative B/P sizes in our sample is also in very
good agreement with the predictions from simulations. For the three simulations
we present in this paper, we find (using the same measurement techniques)
$\rbox/\lbar = 0.40$ for runs A and B and 0.29 for run C. Similarly,
\citet{lutticke00b} reported a relative size of 0.40 from their edge-on analysis
of an $N$-body simulation originally produced by \citet{pfenniger91}. And
\citet{athanassoula02} reported relative B/P sizes (their $LP/L_{2}$) of
0.3--0.6 for a set of three $N$-body simulations, sampled at two different times
each. We can also use the observational results as tests for future theoretical
studies: simulations which produce relative B/P sizes $> 0.6$ will probably not
be good matches to the majority of barred galaxies, though we cannot rule them
out as possible extreme cases.

Finally, the fact that B/P bulges typically span less than half the length of
the bar helps answer objections which have sometimes been raised to the idea
that B/P bulges in edge-on galaxies are due to bars. For example,
\citet{kormendy04} argued that evidence for flat (outer) bars in a few edge-on
galaxies such as NGC~4762, and the fact that boxy bulges have smaller sizes than
bars, presented ``a serious collision between simulations and observations.''
But in reality there is no such collision: both theory and observations agree that
only the inner parts of bars become vertically thickened.

\begin{table}
\caption{B/P and Bar Measurements}
\label{tab:measurements}
\begin{tabular}{@{}lrrrrrr}
\hline
Name & $a(B_{4})$ & $\rbox$  & PA$_{\rm box}$  & $\lbar$ & PA &  $f_{\rm box}$ \\
(1)  & (2)        & (3)       & (4)              & (5)      & (6)&  (7) \\
\hline
\multicolumn{6}{c}{Local Sample} \\
\hline
NGC 1022 & 6.5,9.9 & 8.3 & 140.0 & 19.0 & 22.0 & 0.36 \\ 
NGC 2712 & 11.0,12.0 & 11.0 & 6.8 & 22.0 & 24.0 & 0.35 \\ 
NGC 2962 & 11.0,17.0 & 16.8 & 179.0 & 29.0 & 43.0 & 0.36 \\ 
NGC 3031 & 94.0,108.0 & 97.0 & 149.9 & 130.0 & 210.0 & 0.45 \\ 
NGC 3185 & 15.0,19.0 & 16.0 & 125.3 & 31.0 & 32.0 & 0.47 \\ 
NGC 3368 & 34.0,42.0 & 38.0 & 140.7 & 61.0 & 75.0 & 0.42 \\ 
NGC 3626 & 6.3,12.0 & 11.5 & 162.6 & 20.0 & 26.0 & 0.44 \\ 
NGC 3729 & 11.0,12.0 & 8.8 & 16.4 & 23.0 & 26.0 & 0.31 \\ 
NGC 3941 & 6.7,8.4 & 12.5 & 5.8 & 21.0 & 32.0 & 0.35 \\ 
NGC 4037 & 9.2,12.0 & 8.7 & 11.7 & 27.0 & 33.0 & 0.26 \\ 
NGC 4102 & 6.0,7.5 & 5.1 & 50.7 & 10.0 & 15.0 & 0.29 \\ 
NGC 4143 & 6.1,7.8 & 10.4 & 146.8 & 17.0 & 28.0 & 0.33 \\ 
NGC 4319 & 5.3,10.0 & 7.3 & 152.3 & 15.0 & 17.0 & 0.43 \\ 
NGC 4340 & 27.0,30.0 & 21.0 & 85.9 & 39.0 & 48.0 & 0.27 \\ 
NGC 4386 & 9.6,13.0 & 15.0 & 137.9 & 25.0 & 36.0 & 0.41 \\ 
NGC 4699 & 3.6,4.7 & 6.5 & 45.9 & 13.0 & 16.0 & 0.40 \\ 
NGC 4725 & 40.0,64.0 & 63.0 & 38.5 & 118.0 & 125.0 & 0.50 \\ 
NGC 4995 & 7.1,10.0 & 8.2 & 64.6 & 16.0 & 19.0 & 0.34 \\ 
NGC 5377 & 21.0,39.0 & 26.5 & 37.5 & 58.0 & 67.0 & 0.37 \\ 
NGC 5740 & 5.2,9.5 & 7.4 & 158.2 & 12.0 & 14.0 & 0.36 \\ 
NGC 5750 & 14.0,17.0 & 10.5 & 77.9 & 20.0 & 22.0 & 0.28 \\ 
NGC 5806 & 10.0,20.0 & 16.5 & 169.8 & 37.0 & 38.0 & 0.43 \\ 
NGC 7743 & 13.0,25.0 & 21.5 & 96.6 & 31.0 & 37.0 & 0.58 \\ 
IC 1067  & 5.1,8.1 & 8.5 & 142.0 & 19.0 & 19.0 & 0.43 \\ 

\hline
\multicolumn{6}{c}{Other Galaxies} \\
\hline
NGC 1023 & 22.0,35.0 & 28.0 & 79.7 & 40.0 & 58.0 & 0.41 \\ 
NGC 1808 & 34.0,50.0 & 38.0 & 140.7 & 80.0 & 114.0 & 0.33 \\ 
NGC 2442 & 25.0,38.0 & 29.0 & 55.8 & 57.0 & 65.0 & 0.41 \\ 
NGC 3627 & 17.0,32.0 & 18.0 & 165.9 & 41.0 & 53.0 & 0.32 \\ 
NGC 3992 & 30.0,38.0 & 25.0 & 57.0 & 54.0 & 57.0 & 0.35 \\ 
NGC 4123 & 16.0,27.0 & 15.0 & 112.0 & 48.0 & 50.0 & 0.29 \\ 
NGC 4293 & 17.0,28.0 & 17.0 & 80.0 & 61.0 & 65.0 & 0.27 \\ 
NGC 4535 & 16.0,19.0 & 17.0 & 31.0 & 37.0 & 40.0 & 0.41 \\ 
NGC 5641 & 9.0,13.0 & 11.0 & 159.0 & 23.0 & 24.0 & 0.44 \\ 
NGC 6384 & 8.4,16.0 & 13.0 & 36.5 & 22.0 & 27.0 & 0.48 \\ 
IC 5240  & 13.0,27.0 & 17.0 & 92.6 & 36.0 & 36.0 & 0.46 \\ 

\hline
\end{tabular}

\medskip

Radial lengths and position angles of B/P structures and overall bar
sizes; all lengths are in arc seconds. (1) Galaxy name. (2) Semi-major
axes defined by ellipse-fit $B_{4}$ values: first number = minimum
\bfour{} (= maximum boxyness of isophotes); second = first zero-crossing
of $B_{4}$ outside. (3) Direct measurement of boxy region size on the
image. (4) Position angle of boxy region. (5) Bar semi-major axis. (6)
Position angle of bar. (7) Size of boxy region as fraction of full bar
length $= \rbox/\lbar$ (deprojected).

\end{table}

% figure generated by
% $ ./stats_boxybars_efits.py --histogram-plot=r_box --deltapa-limit=45
% *and*
% $ ./stats_boxybars_efits.py --histogram-plot=r_box
% plus Illustrator (relative_sizes_histogram.ai)

\section{Discussion}

\subsection{Thin Bars: Identifying Galaxies Where the Bar Has Not Buckled}\label{sec:not-buckled}

We have shown that the majority of bars in S0--Sb galaxies probably have B/P
structures, which is consistent with the analysis of edge-on galaxies by
\citet{lutticke00a}. Is this true for \textit{all} bars? The question of whether
\textit{some} bars are indeed flat, without any B/P structure, is an interesting
one. $N$-body simulations generally show that bars undergo a vertical buckling
instability and form B/P structures rather soon (within 1 or 2 Gyr) after the
bar itself forms, and that these structures then persist as long as the bar
does; thus, a barred galaxy without a B/P structure could be an indication of a
very young, recently formed bar.  Alternately, it may be possible to suppress 
buckling  in some galaxies. The buckling instability results from a bar-driven
increase in the in-plane stellar velocity dispersion, which leads to a large
anisotropy in the dispersion \citep{araki85,fridman84, merritt94}. The 
instability can be suppressed, however, if the disc is already vertically hot.
The presence of signficant gas can also suppress buckling, at least in
simulations \citep{berentzen98, debattista06}, while \citet{sotnikova05} suggest
that the presence of a compact, massive, spheroidal bulge could also work.
Finally, the alternative, resonance-trapping mechanism suggested by
\citet{quillen02} predicts that bars should thicken vertically as soon as they
form: in this scenario, \textit{all} bars should have a B/P structure: ``\ldots
barred galaxies should never be found without boxy/peanut-shaped bulges''.

As we have seen, the higher the inclination, the easier it is to detect the
projected signature of a B/P structure -- if the bar's orientation is not too
far away from the major axis of the galaxy (e.g., Figure~\ref{fig:n-body-grid}).
Once the inclination becomes too high (say, $i > 70\degr$), however, it becomes
increasingly hard to directly detect the presence of a bar in the galaxy disc
plane. This is why it is difficult to clearly identify cases in \textit{edge-on}
discs where a bar has formed but has \textit{not} buckled or otherwise thickened
to form a B/P structure.

The best places to look, then, would be barred galaxies with moderately high
inclinations -- e.g., $i \sim 45$--70\degr{} -- where the bar's (deprojected)
position angle is within $\sim 45\degr$ of the major axis. If such galaxies do
\textit{not} show indications of the boxy-bar morphology, then they are good
candidates for systems with completely flat bars.

In Section~\ref{sec:fraction}, we used a slightly more generous limit of $i =
40\degr$ and $\dpa < 45$ when attempting to determine the frequency of the
boxy-bar morphology. Of the 22 galaxies in our local sample meeting those
criteria, we found 14 with at least a boxy interior (and 13 with clear spurs in
addition), which leaves eight galaxies which might lack a B/P structure. If we
increase the inclination limit to 45\degr, then there are six out of 16 galaxies
which do not have good evidence for a projected B/P structure. Most of these are
galaxies with very weak, oval bars and/or evidence for rather luminous bulges,
so that it is more difficult to discern clear morphological features belonging
to just the bars.\footnote{NGC~4941 has slightly boxy isophotes, but no clear
spurs.} However, there are two systems with very strong, narrow bars and no
evidence for large bulges which are our best candidates for barred galaxies
without B/P structures.

\begin{figure*}
\begin{center}
\includegraphics[scale=0.8]{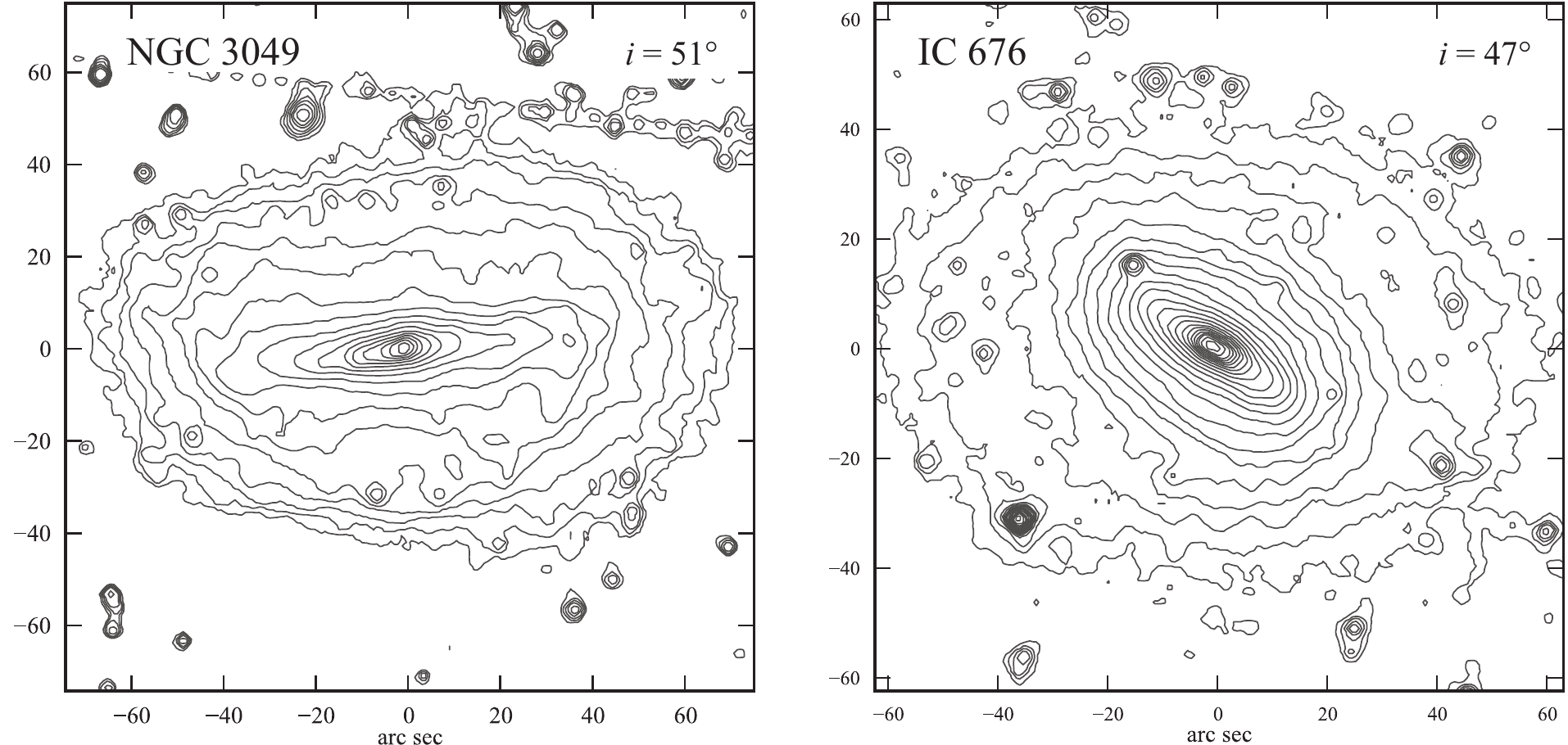}
\end{center}

\caption{\textbf{Left:} Logarithmically scaled isophotes for SBab galaxy
NGC~3049 (inclination $i \approx 51\degr$), using \textit{Spitzer} IRAC1 image from the
SINGS project \citep{kennicutt03}, via NED. \textbf{Right:} Isophotes for the
SB0 galaxy IC~676 ($i \approx 47\degr$), using archival IRAC1 image from S$^{4}$G
\citep{sheth10}. Both images have been rotated to place the disc major axes
horizontal. Despite the relatively high inclinations and favorable bar
orientations (deprojected bar-disc $\dpa = 8\degr$ for NGC~3049, 41\degr{} for
IC~676), there is little or no sign of the box+spurs morphology in either
galaxy, suggesting their bars may not have buckled. Compare with
Figure~\ref{fig:n3049-sim-comparison}.\label{fig:n3049+ic676}}

\end{figure*}

Figure~\ref{fig:n3049+ic676} shows these two galaxies: NGC~3049 and IC~676. They
have inclinations of $\approx 51\degr$ and 47\degr, respectively, and bars
offset from the disc major axis with (deprojected) angles of $\sim 8\degr$ and
41\degr, respectively. Given these orientations and the strength of the bars, we
should be able to see the box+spurs pattern quite clearly.  But as the
figure shows, there is no indication of this: the bars appear to be uniformly
narrow. (A possible hint of narrow, offset spurs is visible in NGC~3049 at a
radius of $\sim 15\arcsec$; however, the apparent offset is in the
\textit{wrong} direction: towards the line of nodes rather than away from it.)
Since there is no evidence for significant bulges in these galaxies -- indeed,
they seem to have little or no bulge at all -- we can rule out the
possibility of a boxy zone being lost within the isophotes of an elliptical
bulge.

\begin{figure*}
\begin{center}
\includegraphics[scale=0.85]{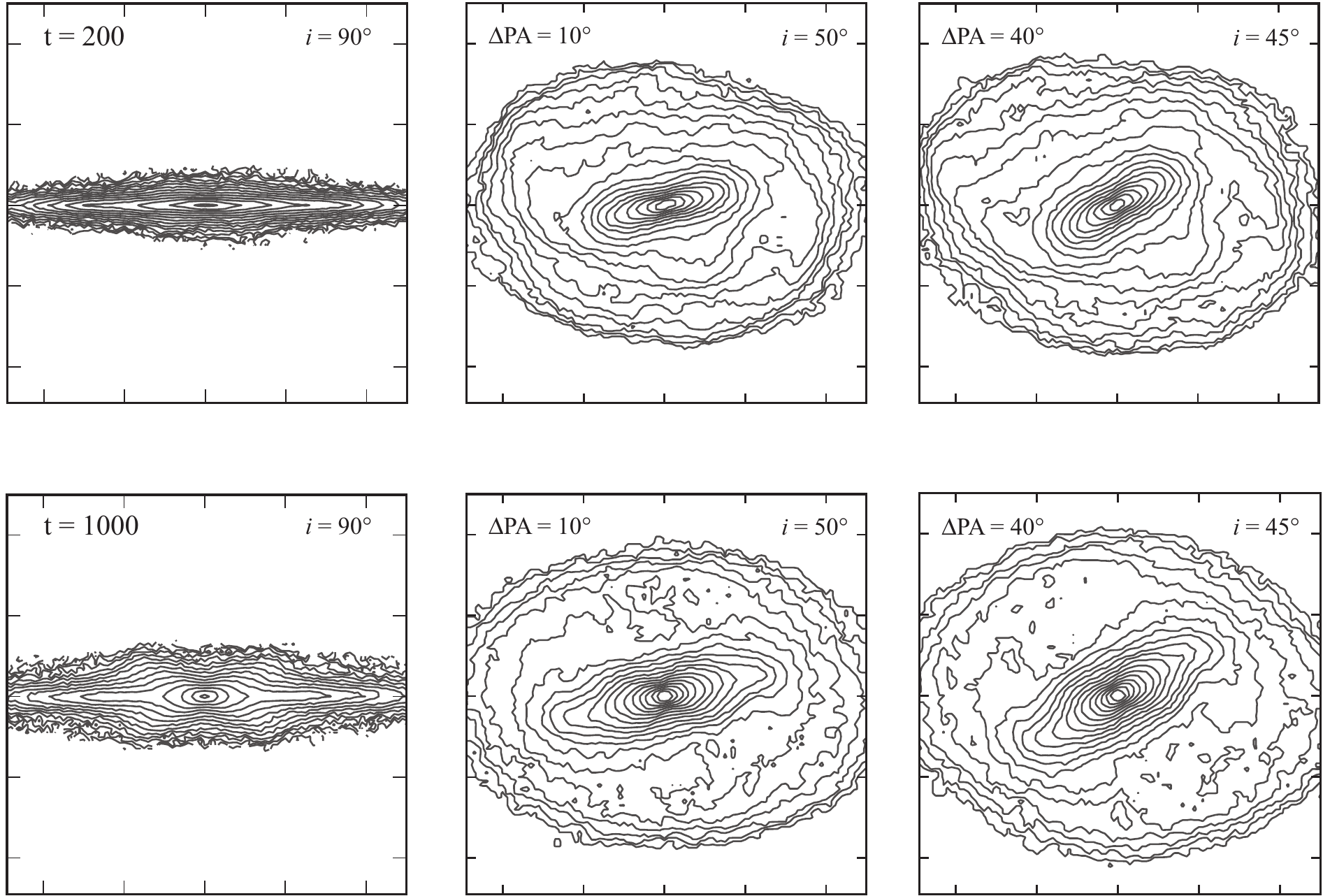}
\end{center}

\caption{Logarithmically scaled isodensity contours from $N$-body model B,
showing two snapshots: $t = 200$ (top panels), prior to bar buckling; and $t =
1000$ (bottom panels), after the bar has buckled. The left-hand panels show the
edge-on ($i = 90\degr$) view with the bar perpendicular to the line of sight, to
emphasize the presence (or absence) of the box/peanut structure; the middle
panels simulate observing the galaxy at an inclination of $50\degr$, with the
bar offset at $\dpa = 10\degr$ (in the disc plane) from the major axis, which is
horizontal; and the right-hand panels do the same but with $i = 45\degr$ and
$\dpa = 40\degr$. In the later snapshot (lower panels), the bar has clearly
buckled, producing strong peanut-shaped isophotes when seen edge-on (lower left)
and a box+spurs morphology, with slightly offset spurs, when seen at
intermediate inclination (lower middle and right). Compare with
Figure~\ref{fig:n3049+ic676}.\label{fig:n3049-sim-comparison}}

\end{figure*}

For comparison, Figure~\ref{fig:n3049-sim-comparison} shows two stages from the
one of our $N$-body simulations: before the bar has buckled (top panels), and
after (bottom panels). When the simulated galaxy is projected with approximately
the same orientations as NGC~3049 and IC~676 (middle and right-hand panels), the top
panels -- showing the simulation \textit{before} bar-buckling -- are clearly 
better matches to the galaxies in question. This agreement suggests that the bars in
NGC~3049 and IC~676 have \textit{not} vertically buckled.

The extremely narrow bars, along with the absence of any sizeable bulge in these galaxies, are
also reminiscent of the nearly face-on ($i = 21\degr$) SBd galaxy NGC~600, where
\citet{mendez-abreu08} failed to find any kinematic signature of a B/P
structure.

Very roughly speaking, then, we can put a lower limit on the frequency of
vertically thin bars at $13^{+11}_{-6}$\%. If we include all of the uncertain
cases -- galaxies with weak, oval bars or large bulges -- then the upper limit
would be $38^{+13}_{-11}$\%. In any case, the existence of thin bars can be used
to help constrain models of B/P structure formation. The resonance trapping
model of \citet{quillen02}, which implies that \textit{all} bars should have B/P
structures, evidently cannot be a very common mechanism. The ability to identify both
buckled and non-buckled bars in larger samples, combined with comparisons of
galaxy properties between buckled and non-buckled bars, will help determine
whether buckling is actually suppressed on long timescales (e.g., when
significant amounts of gas are present, or when discs are vertically hot), or
whether galaxies like NGC~3049 and IC~676 have simply formed their bars recently
enough that buckling has not yet taken place.

\subsection{Things (Mostly) Not Seen: Pinching of the Boxy Zone}

Although the matching of projected bar structure between $N$-body simulations
and actual galaxies with similar orientations can be quite good, there is a
feature of the projected simulations which is rarely seen in real galaxies with
moderate inclinations. Specifically, the isophotes of the boxy zone often show
``pinching'' in the simulations viewed at moderate inclinations (e.g., the $\dpa
= 0\degr$ and 30\degr{} views at $i = 60\degr$ in Figure~\ref{fig:n-body-grid}). The
cause of the pinched isophotes in the simulations is not hard to divine: it is
the signature of a strong peanut structure, something which manifests more
clearly as an ``X'' shape when the simulation is seen edge-on.

Why do real galaxies not show such strong pinching when viewed at moderate
inclinations? The most obvious cause is probably the presence of extra stellar
structure in the inner regions of these galaxies. A compact bulge -- or even a
compact nuclear or inner disc \citep[e.g.,][]{erwin03-id} -- will contribute
rounder isophotes in the central few hundred parsecs, and the resulting summed
isophotes will tend to smooth out the pinching.  Since our $N$-body simulations
were prepared using pure discs (no pre-existing bulges) and do not include any
gas or star formation, this lack of extra, rounder components in the central
regions is not surprising.

\begin{figure}
\begin{center}
\includegraphics[scale=0.75]{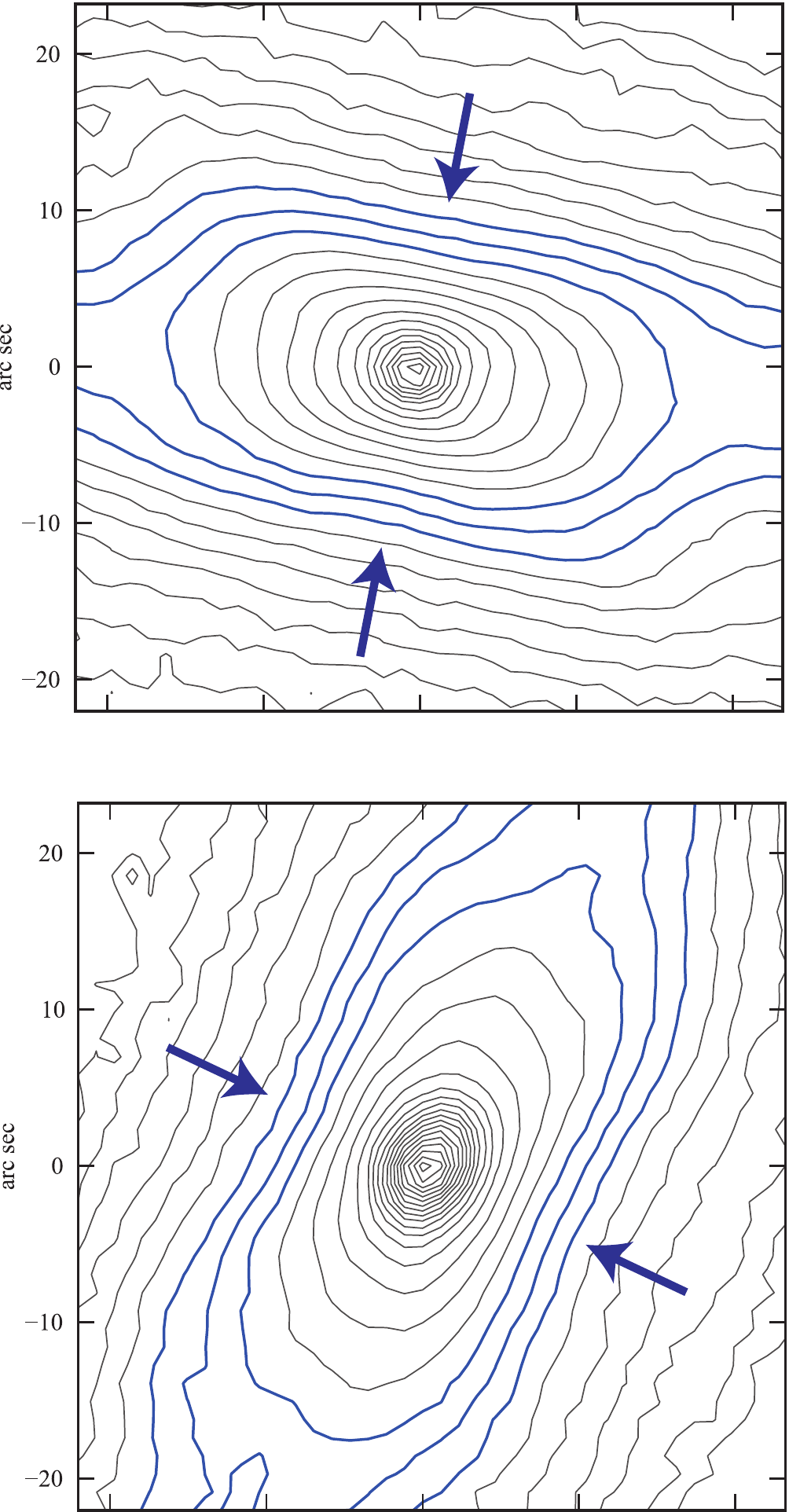}
\end{center}

\caption{Log-scaled $H$-band isophotes for NGC~4293 (top) and NGC~7582 (bottom),
showing evidence for ``pinched'' isophotes in the box region (blue arrows; the
isophotes showing this are outlined with thicker blue lines). Both images are
from \citet{eskridge02}; N is up and E is to the left. \label{fig:pinched}}

\end{figure}

Nonetheless, we can identify \textit{some} real galaxies where boxy zone shows
pinching. Figure~\ref{fig:pinched} shows two such galaxies with $i < 70\degr$. A
very slight hint of pinching can also (perhaps) be seen in the boxy zone of IC~5240's bar
(bottom right panel of Figure~\ref{fig:offset-spurs}).

\subsection{Using the Boxy-Bar Morphology to Constrain Galaxy Orientations}

As we pointed out in Section~\ref{sec:n-body-box-spurs}, the offset spurs in the
box+spurs morphology are due to misalignment between the bar position angle and
the galaxy line of nodes. (If the spurs are symmetric, it means the bar and the
line of nodes have the same position angle.) The projection effects which
produce this also ensure that the visual misalignment between the inner boxy
zone and the spurs is such that the spurs are always offset \textit{away from}
the line of nodes.

This means that it is possible to use the observed boxy-bar morphology to help
distinguish, in a qualitative sense, between possible values of the
galaxy major axis in cases where the latter is uncertain -- e.g., because the
galaxy is warped, interacting, or otherwise strongly asymmetric in its outer
regions.

NGC~2712 is a galaxy in our local sample for which \hi{} mapping by \citet{krumm82}
suggests a kinematic major-axis position angle of $\sim 10\degr$, similar to
that of the (bar-dominated) inner disc. Krumm \& Shane noted that ``beyond about
1\arcmin, however, the optical major axis twists to a position angle $-2\degr$
\ldots This change of position angle is not clearly reflected in the velocity
field, but our poor spatial resolution could hide such an effect.'' The inner
kinematic position angle could be affected by the bar; on the other hand, the
outer optical position angle might be the result of warping or other asymmetry
in the disc. So which position angle better describes the galaxy orientation?

$J$-band isophotes for NGC~2712 can be seen in the upper-right panel of
Figure~\ref{fig:wiyn-galaxies}. The spurs are strongly displaced in a
counter-clockwise direction from the major axis of the boxy region (PA
$\approx 7\degr$, marked by red arrows). If the true line of nodes is at
10\degr, then this morphology is difficult to explain: the spurs should
be offset only slightly, and in the clockwise direction. But if the line
of nodes is instead 178\degr{} (dashed grey line), then the morphology
makes sense: the boxy region is slightly tilted counter-clockwise with
respect to the line of nodes, while the spurs are further offset in the
same direction.

\section*{Summary}

We have presented evidence for a common pattern in moderately inclined barred
galaxies, which we term the ``boxy-bar'' or ``box+spurs'' morphology. In this
morphology, the bar is made of two regions: the interior is broad and slightly
boxy in shape, while the outer part of the bar forms narrower ``spurs''; these
spurs are almost always offset or even rotated with respect to the major axis of
the inner, boxy region.

By comparison with $N$-body simulations, we demonstrate that this morphology
results from the simultaneous projection of the vertically thickened
(``buckled'') inner part of a bar -- the box/peanut (B/P) structure -- and the
vertically thin outer part of the bar. While such structures are often seen in
edge-on galaxies as boxy or peanut-shaped bulges (if the bar is favorably aligned),
we find that they can also be detected for inclinations down to $\sim 40\degr$
-- and, in exceptional cases, as low as $\sim 25$--30\degr.

Examination of ellipse fits to galaxies (real and simulated) with boxy-bar
morphologies shows that a general set of criteria using \afour{} and \bfour{}
(the $\sin 4 \theta$ and $\cos 4 \theta$ deviations from pure ellipticity) exist
for identifying most -- but not all -- cases. However, we argue that ellipse
fits do not provide a consistent and reliable means of measuring the
\textit{size} of the boxy zone, and recommend direct meaasurements on images
instead.

For the latter purpose, we define a visual size measurement for the boxy zone:
\rbox. Comparison of different projections of $N$-body simulations shows that
\rbox{} does an excellent job of describing the radial extent of the B/P
structure as seen in edge-on views; consequently, we are confident that
measurements of \rbox{} in real (moderately inclined) galaxies provides a good
estimate of the extent of B/P structures.

Starting with a local sample of 78 bright S0--Sb barred galaxies with inclinations
$\la 65\degr$, we find 24 galaxies showing some form of the box+spurs
morphology. If we restrict ourselves to the subset of inclinations and relative
position angles (between bar and disc major axis) which maximizes detection of
this morphology, we estimate that at least 2/3 of S0--Sb bars are vertically thickened
in their interiors.

Using the \rbox{} measurement, we find that the B/P structure in our local
galaxy sample spans a range of 0.26--0.58 of the full bar length, with a mean of
$\rbox/\lbar = 0.38 \pm 0.08$; the latter is in excellent agreement with
measurements from a set of six \textit{edge-on} galaxies by \citet{lutticke00b}.
This is clear evidence that when bars thicken vertically, it is only the inner
one- to two-thirds (typically just under half) of the bar which does so.

We note that the combination of being able to easily identify bars when galaxies
are not highly inclined (e.g., $i \la 75\degr$) and the clear features of
projected B/P structures when the galaxy has an inclination $\ga 45\degr$
creates a ``sweet spot'' for finding bars which do \textit{not} have a B/P
structure: galaxy inclination between $\sim 45\degr$ and 70\degr{} and bar
orientation $\la 45\degr$ away from the galaxy line of nodes. From our local
sample, we identify NGC~3049 and IC~676 as plausible candidates for galaxies
with non-buckled (uniformly thin) bars. This implies a lower limit of $\sim
13$\% on the fraction of bars which have not buckled.

\section*{Acknowledgments}

We are happy to thank Inma Martinez-Valpuesta for helpful comments and
conversations, and Ron Buta for suggesting several candidate boxy-bulge
galaxies. Both of them, along with Lia Athanassoula, Alfonso
Aguerri, Dimitri Gadotti, Jairo M{\'e}ndez-Abreu, and Panos Patsis, read
and provided numerous comments on earlier drafts, which helped improve
the paper; we also thank the anonymous referee. P.E. was
supported by the Deutsche Forschungsgemeinschaft through Priority
Programme 1177 ``Galaxy Evolution''; V.P.D. is supported by STFC
Consolidated grant \# ST/J001341/1.

This work is based in part on observations made with the \textit{Spitzer} Space
Telescope, obtained from the NASA/IPAC Infrared Science Archive, both of which
are operated by the Jet Propulsion Laboratory, California Institute of
Technology under a contract with the National Aeronautics and Space
Administration.

Funding for the creation and distribution of the SDSS Archive has been
provided by the Alfred P. Sloan Foundation, the Participating Institutions,
the National Aeronautics and Space Administration, the National Science
Foundation, the U.S. Department of Energy, the Japanese Monbukagakusho, and
the Max Planck Society.  The SDSS Web site is http://www.sdss.org/.

The SDSS is managed by the Astrophysical Research Consortium (ARC) for the
Participating Institutions.  The Participating Institutions are The University
of Chicago, Fermilab, the Institute for Advanced Study, the Japan
Participation Group, The Johns Hopkins University, the Korean Scientist Group,
Los Alamos National Laboratory, the Max-Planck-Institute for Astronomy (MPIA),
the Max-Planck-Institute for Astrophysics (MPA), New Mexico State University,
University of Pittsburgh, University of Portsmouth, Princeton University, the
United States Naval Observatory, and the University of Washington.

Finally, this research has made extensive use of the NASA/IPAC Extragalactic
Database (NED), which is operated by the Jet Propulsion Laboratory, California
Institute of Technology, under contract with the National Aeronautics and
Space Administration.

% References:

\appendix

\section{Plots and Measurements of B/P Structures in Sample Galaxies} %[x]

Figure~\ref{fig:wiyn-galaxies} presents red or near-IR galaxy isophotes for the
24 galaxies in our local sample which display the box+spurs morphology, along
with visual indications of the \rbox{} and \lbar{} measurements for each galaxy;
the numerical values can be found in Table~\ref{tab:measurements}.
Figure~\ref{fig:other-galaxies} does the same for six more galaxies which are
not part of the local sample, taken from Table~\ref{tab:galaxies}; another six
galaxies from Table~\ref{tab:galaxies} can be seen in
Figure~\ref{fig:rbox-demos}.

% wide figure -- use figure*
\begin{figure*}
\begin{center}
\includegraphics[scale=0.89]{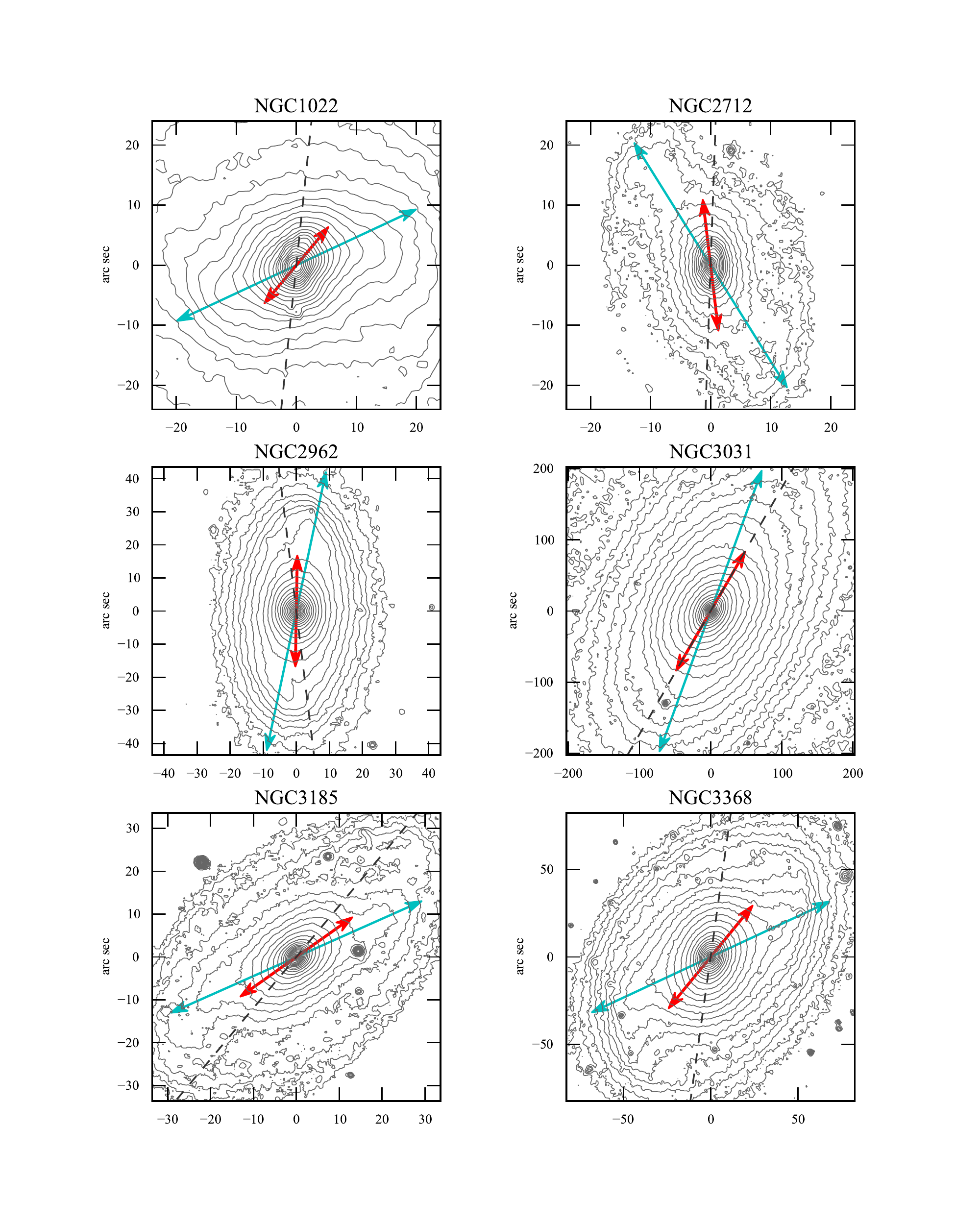}
\end{center}

\caption{Plots of logarithmically scaled isophotes for galaxies in our local
sample with boxy-bar morphology, focussed on the bar region. Dashed black lines
indicates the disc major axis, longer (cyan) arrows indicates position angle and
full length ($2 \times \lbar$) of the bar, and shorter (red) arrows indicate
approximate position angle and full length ($2 \times \rbox$) of the projected
B/P structure. N is up and E is to the left; most isophotes are from near-IR
images (see Appendix~\ref{sec:image-sources} for
details).\label{fig:wiyn-galaxies}}

\end{figure*}

% wide figure -- use figure*
\addtocounter{figure}{-1}
\begin{figure*}
\begin{center}
\includegraphics[scale=0.9]{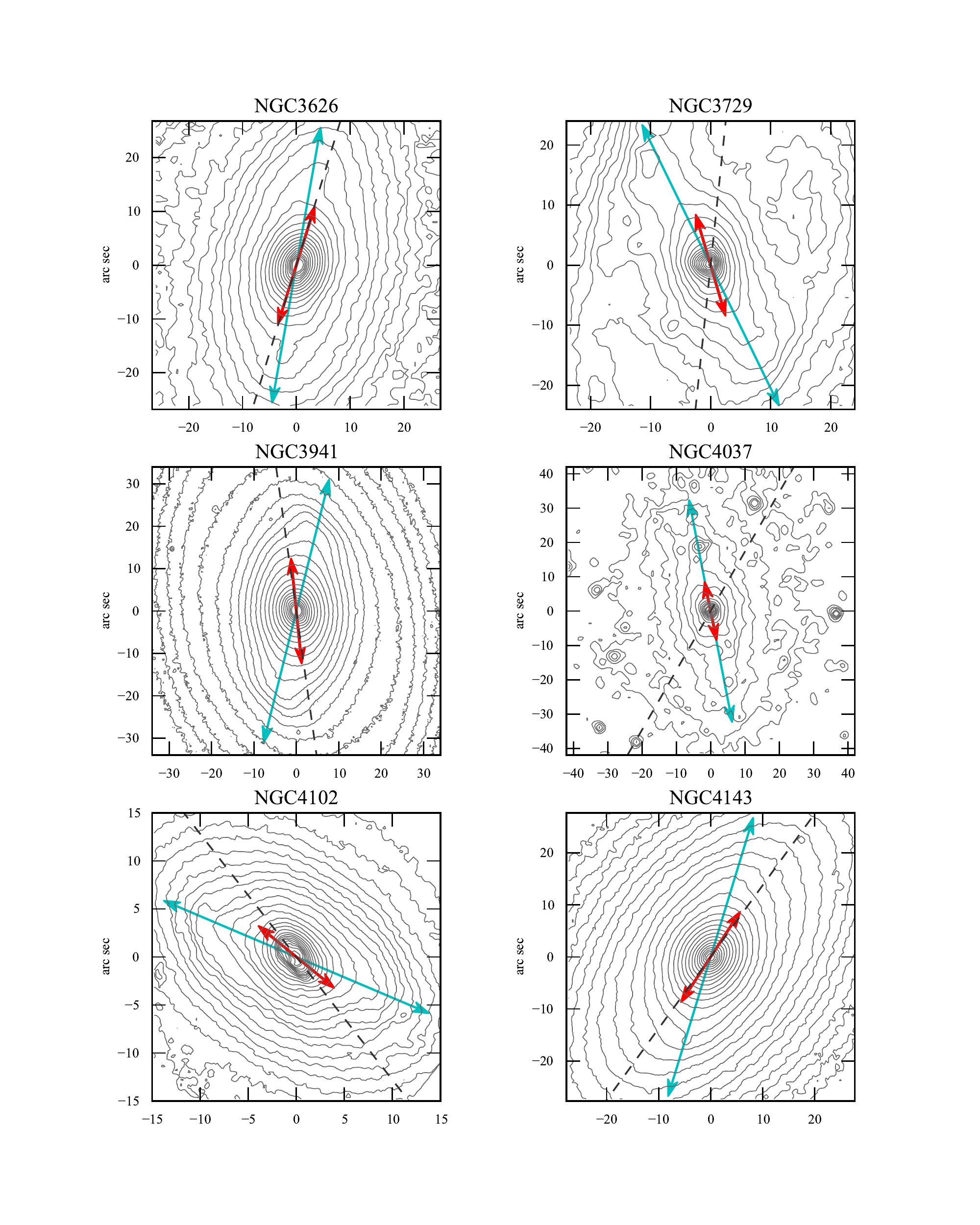}
\end{center}

\caption{-- continued.}

\end{figure*}

% wide figure -- use figure*
\addtocounter{figure}{-1}
\begin{figure*}
\begin{center}
\includegraphics[scale=0.9]{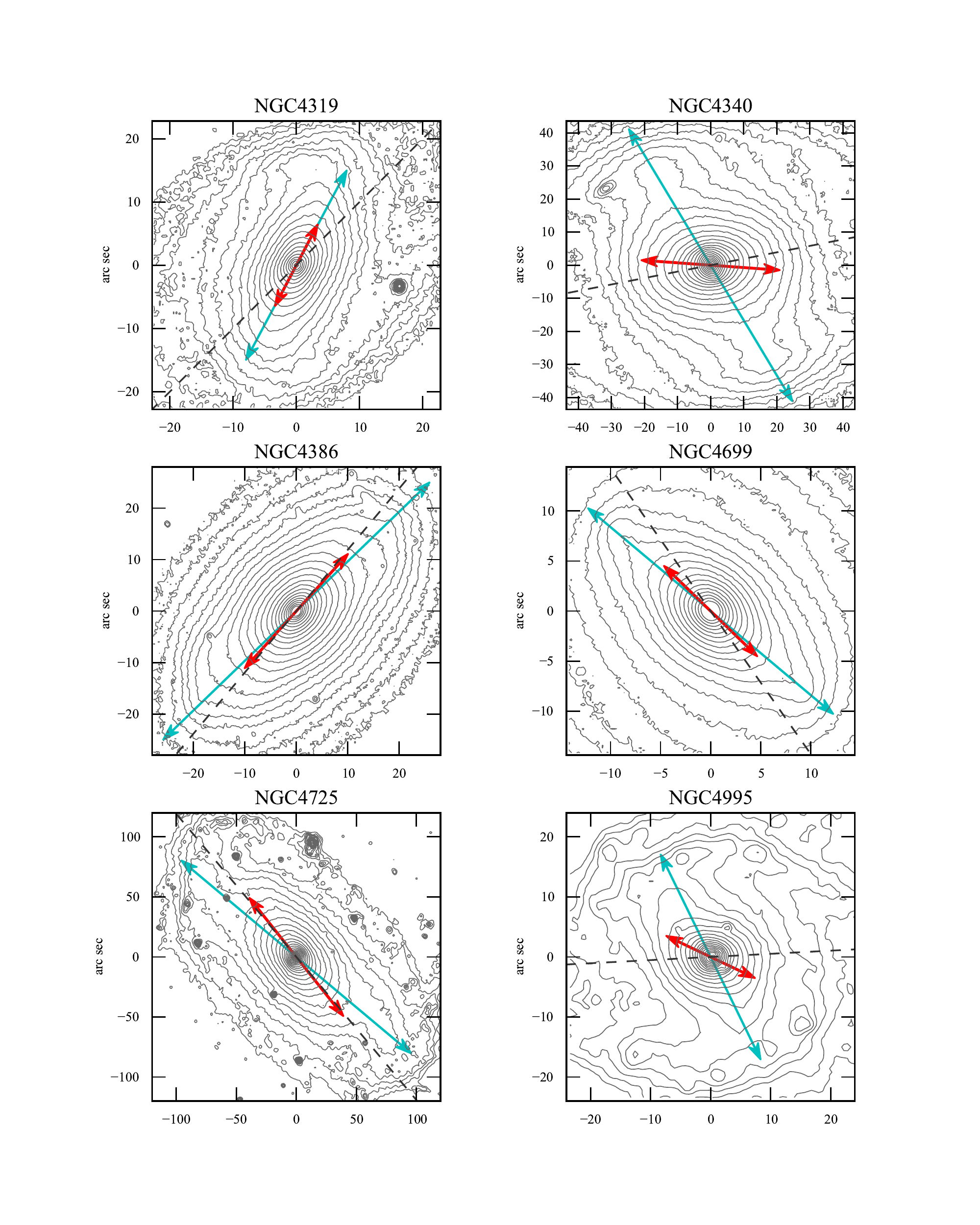}
\end{center}

\caption{-- continued.}

\end{figure*}

% wide figure -- use figure*
\addtocounter{figure}{-1}
\begin{figure*}
\begin{center}
\includegraphics[scale=0.9]{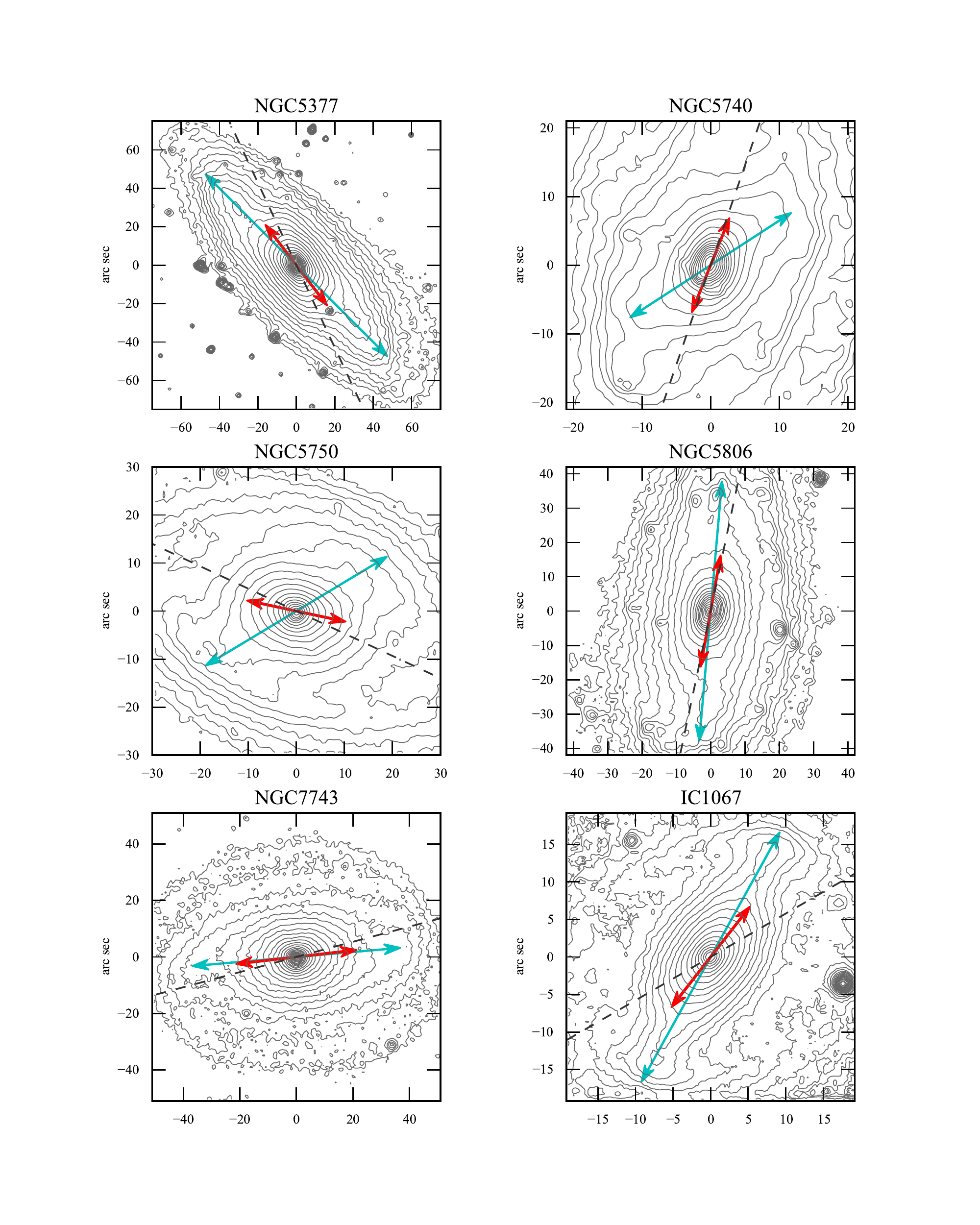}
\end{center}

\caption{-- continued.}

\end{figure*}

% wide figure -- use figure*
\begin{figure*}
\begin{center}
\includegraphics[scale=0.9]{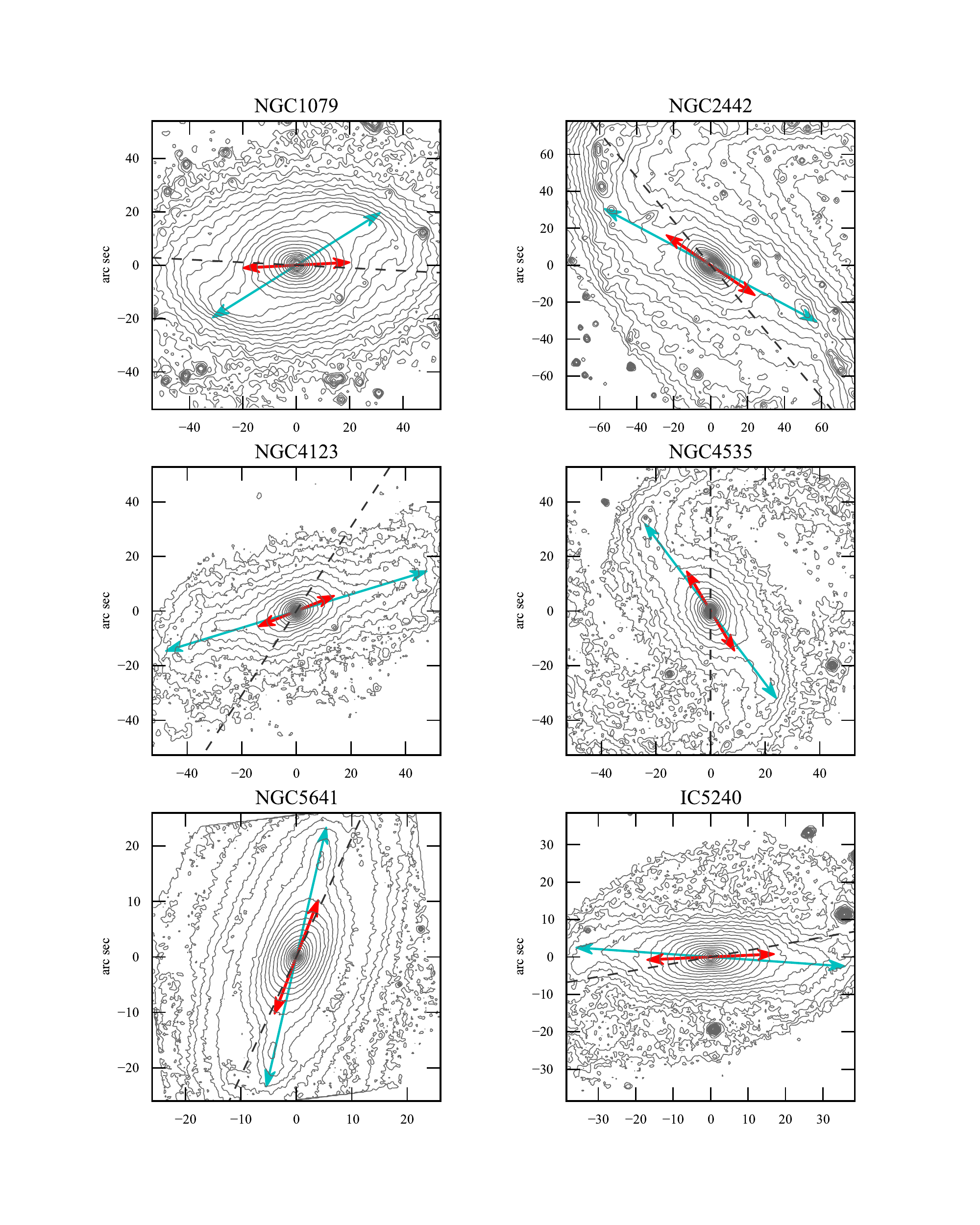}
\end{center}

\caption{As for Figure~\ref{fig:wiyn-galaxies}, but now showing other galaxies with boxy-bar signatures;
see Figure~\ref{fig:rbox-demos} for additional examples.
\label{fig:other-galaxies}}

\end{figure*}

\subsection{Image Sources}\label{sec:image-sources}

We list here the sources and photometric bands of images used in the plots of
the bar regions, including those used for Figures~\ref{fig:wiyn-galaxies} and
\ref{fig:other-galaxies}. Unless otherwise noted, all \textit{Spitzer} IRAC images were
retrieved from the \textit{Spitzer} Heritage Archive; we use the standard post-BCD image
generated by the archive (0.6\arcsec/pixel scale).

\subsubsection{Local Sample}
\textbf{NGC 1022, 4037, 4995, 5740, 5750, 5806:} \textit{Spitzer} IRAC1 images from
\textit{Spitzer} Survey of Stellar Structure in Galaxies \citep[S$^{4}$G;][]{sheth10}.

\textbf{NGC 2712, 4319, 4699, IC 1067:} WHT-INGRID $J$-band images.

\textbf{NGC 2962:} SDSS $i$-band image.

\textbf{NGC 3031, 3368:} \textit{Spitzer} IRAC1 images from \citet{dale09}, via NED.

\textbf{NGC 3185:} WHT-INGRID $H$-band image.

\textbf{NGC 3626:} $K$-band image from \citet{mh01}, via NED.

\textbf{NGC 3729:} \textit{Spitzer} IRAC1 image (Program ID = 61009, PI = W. Freedman).

\textbf{NGC 3941, 4386:} WIYN $R$-band images from \citet{erwin03}.

\textbf{NGC 4102:} \textit{HST} NICMOS3 F160W image from \citet{boker99}, via NED.

\textbf{NGC 4143, 4340:} SDSS $r$-band image.

\textbf{NGC 4725:} \textit{Spitzer} IRAC1 image from SINGS \citep{kennicutt03}, via NED.

\textbf{NGC 5377:} \textit{Spitzer} IRAC1 image (Program ID = 69, PI = G. Fazio).

\textbf{NGC 7743:} \textit{Spitzer} IRAC1 image (Program ID = 40936, PI = G. Rieke).

\subsubsection{Other Galaxies}

\textbf{NGC 1023:} $J$-band image from \citet{mh01}, via NED.

\textbf{NGC 1808, 4293:} $H$-band image from OSU Bright Spiral Galaxy Survey \citep{eskridge02}, via NED.

\textbf{NGC 2442:} \textit{Spitzer} IRAC1 image from \citet{pancoast10}, via NED.

\textbf{NGC 3627:} \textit{Spitzer} IRAC1 image from SINGS \citep{kennicutt03}, via NED.

\textbf{NGC 3992:} \textit{Spitzer} IRAC1 image (Program ID = 80025, PI = L. van Zee).

\textbf{NGC 4123, 4535, 6384:} $K$-band image from \citet{knapen03}, via NED.

\textbf{NGC 5641:} \textit{HST} NICMOS3 F160W image from \citet{boker99}, via NED.

\textbf{IC 5240:} $K$-band image from \citet{mrk97}, via NED.

\end{document}